\documentclass[prb,twocolumn,showpacs,superscriptaddress,aps,longbibliography,citeautoscript]{revtex4-1}  % for review and submission

\usepackage{color}
\usepackage{bm}
\usepackage{bbold}
\usepackage{amsmath}
\usepackage{amssymb}
\usepackage{epsfig,graphicx}
\usepackage{verbatim}
\usepackage[T1]{fontenc}
\usepackage{array}

\newcommand{\be}{\begin{equation}}
\newcommand{\ee}{\end{equation}}
\newcommand{\bw}{\begin{widetext}}
\newcommand{\ew}{\end{widetext}}
\newcommand{\bea}{\begin{eqnarray}}
\newcommand{\eea}{\end{eqnarray}}
\newcommand{\la}{\langle}
\newcommand{\ra}{\rangle}

\newcommand{\p}{\partial}
\newcommand{\rd}{{\rm d}}
\newcommand{\s}{\sigma}
\def\nn{\nonumber\\}
\def\fr#1{(\ref{#1})}

  \usepackage{xspace}
  \usepackage[normalem]{ulem}
  \def\VBSm{VBS$_{-}$\xspace}
  \def\VBSp{VBS$_{+}$\xspace}
  \def\Jx{\ensuremath{J_{\times}}}
  \def\Jp{\ensuremath{J_{\perp}}}

  \newcommand{\aw}[1]{{\color[rgb]{0,0.45,0.74}{#1}}}

  \newcommand{\Fig}[1]{Fig.~\ref{#1}}

  \newcommand{\Eq}[1]{Eq.~\eqref{#1}} % \xspace

  \newcommand{\SU}[1]{\ensuremath{\mathrm{SU}(#1)}}

\usepackage[colorlinks=true,bookmarks=false,citecolor=blue,linkcolor=blue,hyperfootnotes=true,urlcolor=blue]{hyperref}

\makeatletter
\newcommand*{\balancecolsandclearpage}{%
  \close@column@grid
  \clearpage
  \twocolumngrid
}
\makeatother

\begin{document}

\title{Non-Topological Majorana Zero Modes in Inhomogeneous Spin Ladders}

\author{Neil J. Robinson}
\email{n.j.robinson@uva.nl}
\affiliation{Institute for Theoretical Physics, University of Amsterdam, Science Park 904, 1098 XH Amsterdam, The Netherlands} 
\affiliation{CMPMS Division, Brookhaven National Laboratory, Upton, New York 11973, USA}
\author{Alexander Altland}
\affiliation{Institut f\"ur Theoretische Physik, Universit\"at zu K\"oln,
Z\"ulpicher Str. 77, D-50937 K\"oln, Germany}
\author{Reinhold Egger}
\affiliation{Institut f\"ur Theoretische Physik, Heinrich-Heine-Universit\"at, D-40225 D\"usseldorf, Germany}
\author{Niklas M. Gergs}
\affiliation{Institute for Theoretical Physics, Center for Extreme Matter and Emergent Phenomena,
Utrecht University, Leuvenlaan 4, 3584 CE Utrecht, The Netherlands}
\author{Wei Li} % AW: since Wei Li has been out of Munich for many years by now ...
\affiliation{Department of Physics, International Research Institute of Multidisciplinary Science, Beihang University, Beijing 100191, China}
\affiliation{Physics Department, Arnold Sommerfeld Center for Theoretical Physics,
and Center for NanoScience, Ludwig-Maximilians-Universit\"at, 80333 Munich, Germany}
\author{Dirk Schuricht}
\affiliation{Institute for Theoretical Physics, Center for Extreme Matter and Emergent Phenomena,
Utrecht University, Leuvenlaan 4, 3584 CE Utrecht, The Netherlands}
\author{Alexei M. Tsvelik}
\affiliation{CMPMS Division, Brookhaven National Laboratory, Upton, New York 11973, USA}
\author{Andreas Weichselbaum}
\affiliation{CMPMS Division, Brookhaven National Laboratory, Upton, New York 11973, USA}
\affiliation{Physics Department, Arnold Sommerfeld Center for Theoretical Physics,
and Center for NanoScience, Ludwig-Maximilians-Universit\"at, 80333 Munich, Germany}
\author{Robert M. Konik}
\email{rmk@bnl.gov}
\affiliation{CMPMS Division, Brookhaven National Laboratory, Upton, New York 11973, USA}

\date{\today}

\begin{abstract}
{
We show that the coupling of homogeneous Heisenberg spin-1/2 ladders in different phases leads to the formation of interfacial zero energy Majorana bound states. Unlike Majorana bound states at the interfaces of topological quantum wires, these states are void of topological protection and generally susceptible to local perturbations of the host spin system.  However, a key message of our work is that in practice they show a high degree of resilience over wide parameter ranges which may make them interesting candidates for applications.  
}

\end{abstract}

\maketitle
\emph{Introduction:} The Majorana fermion has become one of the most important fundamental quasi particles of  condensed matter physics. Besides its key role as a building block in  correlated quantum matter, much of this interest is motivated by perspectives in quantum information~\cite{RevModPhys.87.137,alicea,alicea2011nonabelian}. Majorana qubits have unique properties which make them ideal candidates for applications in, e.g., stabilizer code quantum computation \cite{sarma2015majorana}. 
Current experimental attempts to isolate and manipulate Majorana bound states (MBSs)  focus on interfaces between distinct phases of symmetry protected topological (SPT) quantum matter. These material platforms have the appealing property that MBSs are protected against local perturbations by principles of topology. In practice, however, topological protection may play a lesser role than one might hope, and various obtrusive aspects of realistic quantum materials appear to challenge the isolation and manipulation of MBSs. Specifically, in topological quantum wires based on the hybrid semiconductor-superconductor platform \cite{Lutchyn2017} or on coupled ferromagnetic atoms \cite{Yazdani2014}, all relevant scales are confined to narrow windows in energy.  In this regard, proposals to realize MBSs in topological insulator nanowires \cite{Cook2011} may offer superior solutions. However, these realizations  require a high level of tuning of external parameters, notably of magnetic fields, and may be met with their own difficulties. 

In this Letter, we suggest an alternative hardware platform for the isolation of zero-energy MBSs. Our proposal does not engage topology. Specifically,  local perturbations  of the microscopic Hamiltonian may induce non-local correlations between the emergent Majorana quantum particles. However, we argue below that in practice this problem is less drastic than one might fear, and that the current architecture  may  grant a high level of effective protection. The numerical evidence provided below certainly points in this direction. 

The material platform we suggest is based on spin ladder materials. Their phases can be classified by combining standard Landau-Ginzburg  symmetry breaking with the presence of SPT order~\cite{wen,*wen1,*wen2,*pollmann,schuch}.
% In SPT phases, symmetry fractionalization occurs on the boundaries~\cite{aklt,wen,*wen1,*wen2,*pollmann}, with associated boundary spin-1/2 degrees of freedom in \SU{2} symmetric ladders. 
We show here that combining ladders in different phases provides a systematic means to generating interface MBSs. The formal bridge between the physics of spin ladders and that of Majorana fermions is provided by a two-step mapping, first representing the spin degrees of freedom by bosons, followed by refermionization of the latter into an effective Majorana theory \cite{Shelton}. We will discuss how numerous spin ladder properties that are difficult to access in the spin language are made simple and transparent in Majorana representation. 
In particular, \SU{2} invariant spin ladders with two legs are described by a theory of four massive Majorana fermions, comprising a triplet and a singlet of different masses, together with a global parity constraint. 
The ground state (g.s.) degeneracies of the spin systems are then encoded entirely in zero-energy MBSs localized on the  boundaries of the system. 

% (Throughout, we tacitly assume that the g.s.~degeneracy equals the degeneracy of the zero energy subspace.)

Two surprising findings arise from this Majorana representation. The first is that additional g.s.~degeneracies can appear in \emph{inhomogeneous} ladders, where the spin-spin interactions vary spatially along the ladder. In the fermionic language, these degeneracies manifest themselves in  new MBSs appearing at the phase boundaries via the Jackiw-Rebbi mechanism \cite{JR}, according to which a sign change in the fermion mass creates a zero mode. This may happen even if all of the bulk phases composing the ladder do not support MBSs on their own.  The second finding is that zero-energy MBSs exist only if the spatial variation of spin couplings about the boundary is sufficiently gentle (a few lattice sites, in practice).
The spatial smoothness across the interface is required to stabilize the mapping onto a continuum description and to prevent the coupling of distant MBSs via higher-energy states.
This condition manifests the lack of topological protection. (For other zero energy modes in topologically trivial phases, see~Refs.~\onlinecite{Yan2017,Chan2017,Yan2018,topo_bootstrap,*topo_boot_kondo,simon}.)
However, we present numerical evidence that these MBSs are nonetheless close to zero energy over parametrically wide regions.

\begin{figure}[t]
\includegraphics[width=0.425\textwidth,trim={18 45 35 93},clip]{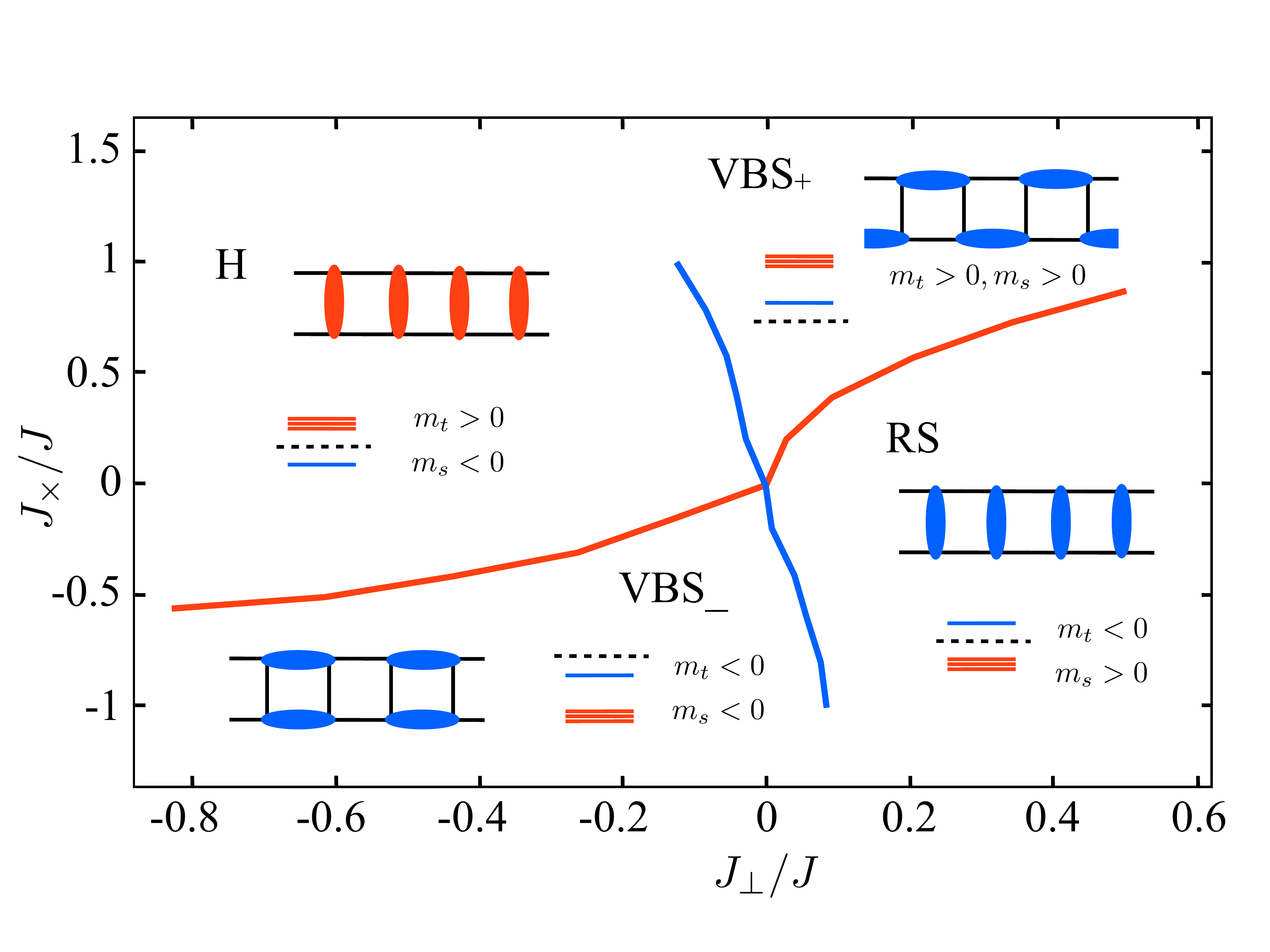}
\caption{Phase diagram for the Hamiltonian~(\ref{SpinHam}) obtained from
SU(2) DMRG simulations of a $100\times2$ site ladder with bond dimension $\chi = 1500$ states. The red (blue) phase boundary shows the critical line with Majorana fermion mass $m_t=0$ $(m_s=0$). 
Inset figures show schematic representations of singlet (blue) and triplet (red) bond order 
within each phase, and the corresponding signs of $m_t$ and $m_s$.}
\label{Fig:PhaseDiag}
\end{figure}

 \emph{Spin ladders:} Ladder geometries provide an important viewport on the physics of strongly correlated electron systems~\cite{Dagotto618} and are a research focus of condensed matter physics in their own regard.  They are close enough to being one-dimensional (1D) that  powerful theoretical techniques can be deployed in their understanding running the gamut from field theory~\cite{Shelton,PhysRevB.91.174407,PhysRevB.64.155112,PhysRevLett.96.086407} and Bethe ansatz~\cite{PhysRevB.60.9236,*batchelor} to density matrix renormalization group (DMRG)~\cite{White92,Schollwoeck05,Schollwoeck11,NOACK1996281,PhysRevB.89.094424,weichselbaum2018unified}. However, they are also far enough removed from 1D that they capture the physics of  two-dimensional systems.   We here focus  on ladders where the fluctuations of spin-1/2 degrees of freedom are dominant (e.g., SrCu$_2$O$_3$~\cite{PhysRevLett.73.3463}) over ladders where charge degrees are mobile (e.g., Sr$_{14-x}$Ca$_x$Cu$_{24}$O$_{41}$~\cite{VULETIC2006169}). For concreteness, we consider the two-leg ladder Hamiltonian
\be
\begin{split}
H =& \ J \sum_{\ell=1,2}\sum_{r=1}^{N-1} \bm{S}_{\ell,r}\cdot\bm{S}_{\ell,r+1} + \Jp \sum_{r=1}^N \bm{S}_{1,r}\cdot\bm{S}_{2,r} \\
& + \Jx \sum_{r=1}^{N-1} \Big( \bm{S}_{1,r}\cdot\bm{S}_{1,r+1} \Big)\Big( \bm{S}_{2,r}\cdot\bm{S}_{2,r+1}\Big), 
\end{split}
\label{SpinHam}
\ee
where $S^a_{\ell,r}$ is the $a=x,y,z$ spin-1/2 operator located on leg $\ell$ and rung $r$ of the ladder. The exchange parameters $J\aw{:=1},\,\Jp,\,\Jx$ characterize leg, rung, and plaquette interactions, respectively.  For  uncoupled Heisenberg chains, the total spin of each leg would be conserved, and we could work in a representation where 
$S^z_\ell=\sum_r S^z_{\ell,r}$ are good quantum numbers. Assuming an even number $N$ of sites per chain, both $S^z_{\ell}\in \Bbb{Z}$ are integer valued. The coupling $J_\perp$  exchanges spin  in integer units, $S^z_1\to S^z_1\pm 1, S^z_2\to S^z_2\mp 1$, violating the conservation of the individual $S^z_{\ell}$, but still constraining the even and odd combinations,  $S^z_\pm=S^z_1\pm S^z_2$, to have identical \textit{parity},
\begin{equation}\label{paritycond}
S^z_+ \equiv S^z_- (\rm mod 2).
\end{equation}
We thus expect an effective fermionized theory of the system to display a $\mathrm{U}(1)$ symmetry reflecting the conservation of $S^z_+$ plus a  $\mathbb{Z}_2$ parity condition implementing \eqref{paritycond}. The latter introduces correlation between the $S^z_+$ and the $S^z_-$ sector and will play a key role throughout.

\begin{table}[t]
\renewcommand{\arraystretch}{1}% Spread rows out...
\begin{tabular}{>{\centering\arraybackslash}m{2cm} >{\centering\arraybackslash}m{2cm} >{\centering\arraybackslash}m{2cm} >{\centering\arraybackslash}m{2cm}}
\hline
\hline
 Phase &  $m_t/m_s$  & $d_+/d_-$ & g.s.~deg.\\
 \hline
 H & $+/-$ & 4/2 & 4 \\
 RS  & $-/+$ & 1/2 & 1 \\
 \VBSp & $+/+$ & 4/4 & 8\\
 \VBSm & $-/-$ & 1/1 & 1 \\
 \hline
 \hline
\end{tabular}
\caption{Phases of the spin model, the signs of their fermion masses, $m_t/m_s$, the g.s.~degeneracies, $d_{\pm}$, of their even/odd sectors ($S^z_{\pm}$) before the parity restriction \eqref{paritycond} is applied, and finally their overall actual g.s.~degeneracies from \aw{SU(2)} DMRG.}
\label{Tab:Comparison}
\end{table}

\emph{Phase diagram:} Depending on the couplings $J_\perp, J_\times$, the Hamiltonian~\eqref{SpinHam} supports different phases. For strong positive rung interaction $J_\perp$ and weak plaquette interaction $J_\times$, the formation of rung singlets (RS) is favored, cf.~the lower right part of Fig.~\ref{Fig:PhaseDiag}. For strong negative couplings $J_\perp$, rung triplets are formed instead and effectively implement an $S=1$ Haldane-Heisenberg chain (Haldane phase, H). For strong  $J_\times$, one may anticipate `valence bond solids' (VBS) distinguished by different types of periodically repeated intra-chain dimerization, \VBSp and \VBSm (see Fig.~\ref{Fig:PhaseDiag}). 
While the existence of different dimerization patterns is relatively easy to anticipate, it takes more effort  to determine the symmetries characterizing them, the respective order parameters, the g.s.~degeneracies, and the phase boundaries. For example, the Haldane phase is an SPT phase without a local order parameter. It exhibits a four-fold g.s.~degeneracy due to two spin-$1/2$ degrees of freedom dangling at the boundaries. In particular, the identification of the symmetries of the VBS phases is a non-trivial matter \cite{fuji1,fuji2}. The boundaries between the phases as well as the ensuing g.s.~degeneracies can be established via DMRG simulations (see Appendix D): in Fig.~\ref{Fig:PhaseDiag} we present the phase diagram and in Tab.~\ref{Tab:Comparison} the g.s.~degeneracies.

\begin{figure}
\includegraphics[width=0.425\textwidth]{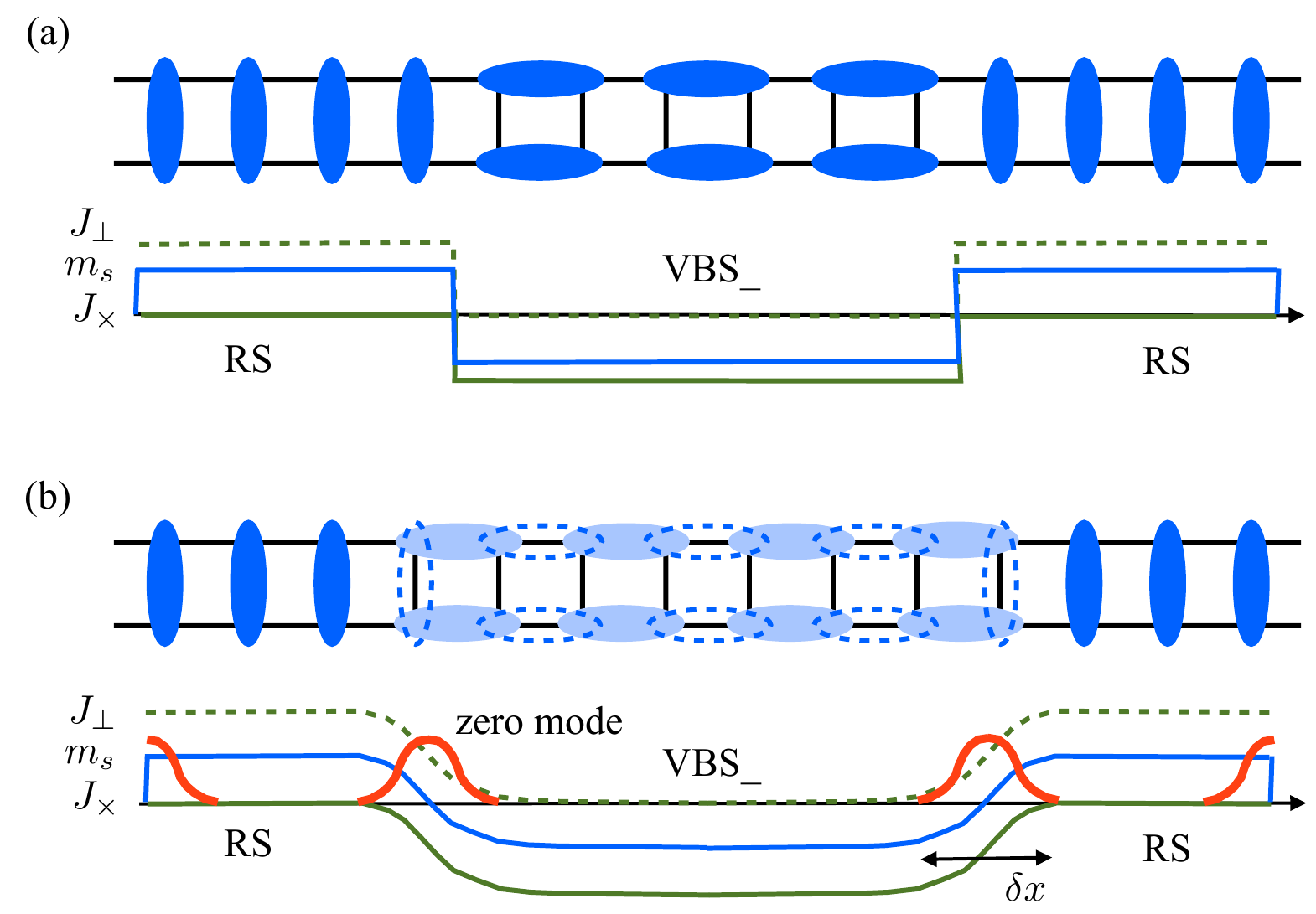}
\caption{
Bond formation patterns when parameters $(\Jx,\Jp)$ are varied (in green) in order to form a RS--\VBSm--RS ladder.
a) For sudden parameter changes at interfaces. 
b) For smooth parameter changes, Jackiw-Rebbi zero modes emerge when the singlet mass, $m_s$, changes sign (lowest sketch).   }
\label{Fig:RSVBSRSsketch}
\end{figure}

The presence of distinct dimerization patterns also provides a first clue as to the formation of zero 
energy degrees of freedom if chains of competing order are coupled by interfaces of sufficiently smooth variation. As an example, consider the RS--\VBSm--RS setup in Fig.~\ref{Fig:RSVBSRSsketch}. The \VBSm chain breaks a translational $\mathbb{Z}_2$ symmetry via the choice of the links harboring singlet configurations (indicated as blue ovals). If the interface is sharp, one such configuration is rigidly pinned between two RS phases, and the ground state is unique. However, for a smooth interface, dimerization patterns of either parity can be put at no difference in energy (cf. the bottom part of the figure). This leads to a $\mathbb{Z}_2$ g.s.~degeneracy between phases whose ground states are individually non-degenerate.

\emph{Majorana representation:} All the structures and phenomena alluded to above afford a simple and surprisingly  quantitative description in a language of Majorana fermions. The passage to this representation involves the abelian bosonization \cite{EA1992} of the spin ladder as an intermediate step. In a second step, the bosonic degrees of freedom are mapped to an equivalent system of Majorana fermions \cite{Shelton}. Within the bosonized framework, smooth and rapid changes of the spin magnetization in the interaction terms are represented as gradient (`current-current') and transcendental (`massive') perturbations of the boson fields, respectively (see Appendix A.1). Within the fermion language, these in turn correspond to interaction terms and bilinear fermion operators, where, crucially, the former turn out to be irrelevant in a renormalization group sense. This means that, perhaps counter-intuitively, the spin ladder is represented by a system of two \textit{non-interacting} fermion fields, representing the sum and the difference $S_\pm$ of the magnetization, respectively. The fermion bilinears describe scattering between left and right moving fermions, plus effectively superconducting correlations in the $S^z_-$ sector reflecting the absence of $\mathrm{U}(1)$ symmetry. Much as for the case of topological superconducting wires \cite{alicea}, it then pays off to switch to a Majorana fermion representation. As a result, one arrives at the low-energy continuum Hamiltonian 
\be
\begin{split}
H = \int &\rd x \bigg[-\frac{iv}{2} \Big( \xi^0_R \p_x \xi^0_R - \xi^0_L \p_x \xi^0_L \Big) - i m_s \xi^0_R \xi^0_L \\
&- \frac{iv}{2} \Big( \bm{\xi}_R \p_x \bm{\xi}_R - \bm{\xi}_L \p_x \bm{\xi}_L \Big) - i m_t \bm{\xi}_R\cdot \bm{\xi}_L\bigg],
\end{split}
\label{MajHam}
\ee
where $\xi^{0,1,2,3}$ are Majorana fields
 arranged into a singlet, $\xi^0$, and a triplet, $\bm{\xi} = (\xi^1,\xi^2,\xi^3)$, subject to masses \cite{Shelton}
\be
m_t \propto 9\Jx/\pi^2 - \Jp, \qquad
m_s \propto 9\Jx/\pi^2 + 3\Jp.
\label{msmt}
\ee
 The doublets $(\xi^1,\xi^2)$ and $(\xi^0,\xi^3)$ represent the $S^z_+$ and $S^z_-$  sectors, respectively. In the Majorana language, the $\mathrm{U}(1)\simeq\mathrm{O}(2)$ symmetry of the 
$S^z_+$ sector is realized as a continuous rotation symmetry between the mass-degenerate fields $(\xi^1,\xi^2)$, and the $\mathbb{Z}_2$ symmetry of the $S^z_-$ sector via sign inversion of $\xi^{0,3}$. Importantly, these Majorana fields are not independent but  correlated via the spin parity relation \eqref{paritycond}. In the present language, the global $S^z_\pm$ quantum numbers assume the form $S^z_+=i\sum_a \xi_a^2 \xi_a^1/2$ and $S^z_-=i\sum_b \xi_b^3\xi_b^0/2$, where $\sum_{a,b}$ is a formal sum over all eigenmodes of the system. (In translational invariant cases, these are momentum modes. However, for systems with boundaries or interfaces, the situation gets more interesting.) The constraint \eqref{paritycond} thus translates to %the condition
\begin{equation}\label{parity2}
    \exp(\pi\sum_a \xi^1_a\xi^2_a/2)=\exp(\pi\sum_b \xi^3_b\xi^0_b/2),
\end{equation}
introducing entanglement between the four Majorana sectors (see Appendx A.2).

\emph{Interfacial Majorana states:} In the Majorana representation, the g.s.~degeneracy of a phase is diagnosed via the appearance of MBSs localized at the system's boundaries. Here the vacuum can be represented as a fictitious Majorana system with infinitely large negative mass \cite{Lecheminant}. A vacuum interface of a system with bulk positive mass then amounts to the zero-crossing of a spatially dependent mass function $m(x)$, where the Jackiw-Rebbi mechanism implies the presence of a zero-energy MBS at each end. Since two MBSs define a fermion Hilbert space of dimension two, prior to imposing the parity constraint \eqref{parity2}, the g.s.~degeneracy of a system of definite $(J_\times,J_\perp)$ is given by $d=d_+d_-$, $d_+= 2^{2\Theta(m_t)}$, $d_-=2^{\Theta(m_t)+\Theta(m_s)}$, where $\Theta$ is the Heaviside function and we use Eq.~\eqref{msmt}. For $d>1$,~\eqref{parity2} then implies a factor of two reduction in the 
actually realized g.s.~degeneracy, $d\to d/2$.  This integer agrees exactly with the DMRG results listed in Tab.~\ref{Tab:Comparison}. The same g.s.~degeneracies also follow from the bosonized formulation (see Appendix B) from a truncated conformal space approach \cite{YZ1,*YZ2,review} for sine-Gordon like models~\cite{feverati1998truncated,feverati1998scaling,feverati1999non,bajnok2004susy,bajnok2002finite,bajnok2002spectrum,bajnok2001nonperturbative,bajnok2000k,takacs2006double,toth2004nonperturbative,palmai2013diagonal,scnt,scnt1,BajnokNuclPhysB01,BajnokNuclPhysB02}.

\begin{figure}
\includegraphics[width=0.8\linewidth]{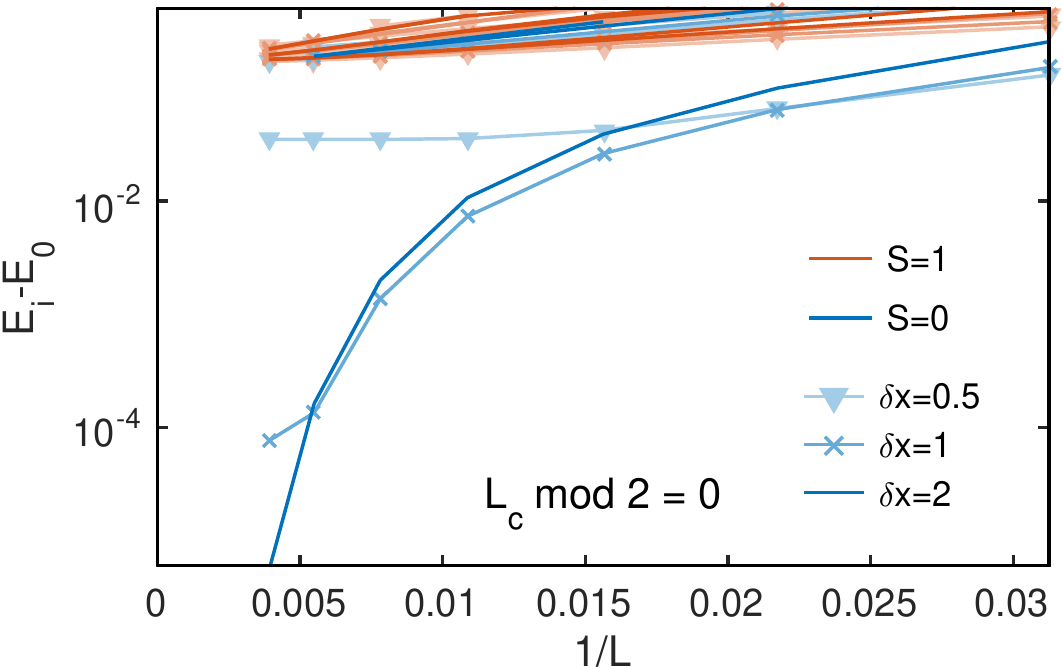}
\caption{Finite size scaling of low-energy  SU(2) DMRG eigenstates in RS--\VBSm--RS ladders of total length $L$. 
  Blue/red indicates singlet/triplet states ($S=0/1$) (the g.s. is not shown).  We use $J=1$, with the other couplings
  varied as $\Jp(x) = \tfrac{4}{3} (1-w(x))$ and $\Jx(x) =-\tfrac{4}{3} w(x)$, with $w(x)=f(x-x_+)  - f(x-x_-)$,   $f(x)=[1+\exp(\tfrac{x}{\delta x})]^{-1}$,   $L_c \equiv x_+-x_-$, and $x_\pm=(L\pm L_c)/2$.  The width $\delta x$ controls the interface smoothness.  The g.s.~degeneracy develops quickly with increasing $\delta x$ and is only marginally affected by the length
  $L_c$ of the \VBSm region (see Appendix D).
}
\label{Fig:RS-VBSm-RS}
\end{figure}

What happens at interfaces between ladders of different symmetry can now be understood in equally straightforward ways. Let us then return to the RS--\VBSm--RS hybrid, see Fig.~\ref{Fig:RSVBSRSsketch}. Provided the interface varies smoothly enough, the system is described by the Majorana theory with $m_t<0$ but $m_s$ changing from positive values to negative and back. We thus have MBSs at both interfaces with spatial extension determined by the width of the interface region. Naively, one might think that the same principle secures the existence of MBSs in the complementary case of \VBSm--RS--\VBSm hybrids as well. However, there is a catch: The above argument does not make reference to the parity constraint \eqref{parity2}. In the RS--\VBSm--RS case, since $m_s>0$ in the outer RS segments, MBSs will not only exist at the internal interfaces but also at the outer vacuum boundaries, cf.~Fig.~\ref{Fig:RSVBSRSsketch}(b). This implies that changes in the occupation of the internal MBS system can be compensated by the outer MBS system, which may act as a 'parity sink' to restore the condition \eqref{parity2}.
In concrete terms, the $+$ sector of the RS--\VBSm--RS ladder is even parity and has a unique g.s. as $d_+=1$ for the RS and \VBSm segments.  On the other hand, the $-$ sector is nominally 4-fold degenerate (as $d_-=2$ for each RS segment), but only two of the four states have even parity, thus leaving only two allowed states once we combine the $\pm$ sectors.

In Fig.~\ref{Fig:RS-VBSm-RS} we present DMRG results showing that the RS--\VBSm--RS ladder indeed has a doubly degenerate ground state for smooth interfaces. If $\Jp$ and $\Jx$ defining these phases vary too sharply, the ground state remains unique. We explain why this is so field theoretically in Appendix C.  However, once the scale of variation extends over just a few lattice sites, one rapidly approaches a two-fold degenerate g.s. We also note that the energy gap protecting the g.s.~degeneracy is rather large for the example in Fig.~\ref{Fig:RS-VBSm-RS}.   
%It is remarkable that in the RS--\VBSm--RS example, MBSs are generated in a system none of whose individual parts, \VBSm or RS, supports such states. 
It is remarkable that MBSs are generated in the RS--\VBSm--RS example, where none of the individual parts, \VBSm or RS, support such states. Those MBSs also provide a means to distinguish two different SPT-trivial phases, cf.~Refs.~\onlinecite{fuji1,fuji2}. The situation is rather different for the  \VBSm--RS--\VBSm system. Since one of the two fermion states formed from the central MBS pair is parity blocked, MBSs are effectively removed from the zero energy Hilbert space \footnote{One may change the occupation of the MBS pair only at the expense of populating high-energy states via a mechanism similar to the 'quasiparticle poisoning' \cite{alicea} of topological Majorana wires.}. See Appendix D.2.b for verification of this via DMRG.  In this way, the parity constraint trumps the Jackiw-Rebbi principle.

Interfaces between phases of enriched symmetry define higher-dimensional MBS systems. As an example, consider the RS--H--RS hybrid. Although the g.s.~degeneracy of the central H segment (the outer RS phases) is only four-fold (unique), 
the interfaces harbor a potential 32D zero-energy space, with four MBSs at either side of the H segment since four masses change sign upon crossing from one phase into the other. Parity, as in the RS--\VBSm--RS ladder, reduces this by one-half (see Appendix D.2.c).
%the interfaces harbor a 16-dimensional zero-energy space, %with four MBSs at either side of the H segment since four %masses change sign upon crossing from one phase into the %other. Parity is not an issue here, as the two MBSs at the %outer ends act as parity sinks.

\emph{Reality check:} The above constructions demonstrate that spin ladder materials provide a remarkably rich platform for the isolation of zero energy MBSs, with sizeable energy gaps to higher-lying states. In view of the general interest in MBSs it is imperative to ask how our non-topological MBSs fare in comparison to topologically protected MBSs. At first sight, the absence of topological protection appears to be a crucial setback. However, 
at present the probably most obtrusive effect hampering Majorana device functionality is the buildup of long-range MBS hybridizations. In topological devices the hybridization exponentially approaches zero with increasing distance but can nonetheless be large in practice. For example, in hybrid semiconductor wires, topological protection crucially relies on the rather tiny superconducting proximity gaps  \cite{alicea,mourik2012signatures,chang2015hard,higginbotham2015parity,albrecht2016exponential}. In the present setup, the lack of topological protection manifests itself in long-range correlations between MBSs 
when short range correlations of the underlying spin chains are changed (in particular, the interface roughness). However, the degrees of freedom behind such changes are highly inert in realistic systems since they 
require energy scales comparable to the exchange couplings. Even though these energy scales do not grow with system size, they can be sufficiently large to provide efficient MBS protection at low temperatures. 

\emph{Outlook:} A promising aspect of our approach is that it brings a plethora of material platforms into play. While we have focused on spin ladders, similar considerations apply to many quasi-1D materials, in particular those that admit a bosonization treatment, e.g., $N$-leg Heisenberg ladders with \SU{2} spin symmetry \cite{Dagotto618,PhysRevB.58.6241,PhysRevB.89.094424} or a more general \SU{M} symmetry \cite{PhysRevB.93.134415,PhysRevB.91.174407,weichselbaum2018unified},  coupled chains of 
itinerant electrons \cite{PhysRevLett.96.086407,PhysRevB.65.115117,loss,smitha}, or coupled Luttinger liquid systems \cite{PhysRevB.89.085101,*PhysRevB.95.125130,JIANG2018}.
In addition, our setup directly comes with an intrinsic source of strong entanglement. Indeed, the Majorana parity constraint \eqref{parity2} plays a similar role to the strong Coulomb charging energy \cite{Fu2010,Beri2012,Altland2013,Beri2013} in mesoscopic MBS systems, where a related parity constraint implies qubit functionality \cite{Plugge2017,Karzig2017}. 
The question of how this entanglement mechanism may be turned into an operational resource, and how the MBSs discussed here can be probed and/or manipulated, is an interesting subject for future study.  

\begin{acknowledgements}
N.J.R. thanks F. Burnell, F. Harper, and D. Schimmel for discussions. Work at BNL (N.J.R., A.M.T., A.W., R.M.K.) was supported by the CMPMS Division funded by the U.S. Department of Energy, Office of Basic Energy Sciences, under Contract No. DE-SC0012704. N.J.R. was supported by the EU Horizon 2020 program, grant agreement No.~745944.  A.W. acknowledges support from the Deutsche Forschungsgemeinschaft (DFG), Grant Nos.~ WE4819/2-1 and WE4819/3-1. D.S. is member of the D-ITP consortium of the Netherlands Organisation for Scientific Research.  
A.A.~and R.E.~acknowledge DFG support via Grant No. EG 96/11-1 and CRC TR 183 (project C4).   R.M.K. and A.M.T. acknowledge the hospitality of LMU Munich and of HHU D\"usseldorf where parts of this work have been done.
\end{acknowledgements}

\appendix

%%%%%%%%%%%%%%%%%%%%%%%%%%%
\section{Majorana representation of the spin Hamiltonian
\label{app:Majrep}}
%%%%%%%%%%%%%%%%%%%%%%%%%%%

\subsection{Abelian bosonization} % (fold)
\label{sub:abelian_bosonization}

% subsection abelian_bosonization (end)
Referring to Refs.~\onlinecite{EA1992,Shelton,Lecheminant} for details, we here
review how the ladder Hamiltonian, Eq.(1) of the main text, is bosonized. Consider the spin operator ${\bm S}_{\ell,r}$ at the point $r = x/a_0$ of the $\ell$th leg, where the lattice spacing is $a_0$. The abelian bosonized description involves splitting the operator into a smooth and staggered component, with these components expressed in terms of a bosonic field $\Phi_\ell$ together with its dual $\Theta_\ell$~\cite{EA1992,Shelton,Lecheminant},
\be
\begin{split}
\frac{S^z_{\ell,r}}{a_0} &= - \frac{1}{2\sqrt{2}\pi}\partial_x \Phi_\ell (x) + \frac{\lambda (-1)^r}{2\pi a_0}\sin \bigg(\frac{\Phi_\ell(x)}{\sqrt{2}}\bigg), \\
\frac{S^\pm_{\ell,r}}{a_0} &= \frac{\lambda e^{\mp i \Theta_\ell(x)/\sqrt{2}}}{2\pi a_0} \bigg[\cos\bigg(\frac{\Phi_\ell(x)}{\sqrt{2}}\bigg) + (-1)^r\bigg].
\end{split}
\label{SpinOp}
\ee
Here $\lambda$ is a non-universal constant related to the frozen charge degrees of freedom of a parent Hubbard ladder \cite{EA1992,Shelton,Lecheminant}. The Hilbert space of each boson is divided into sectors marked by the total $S^z$ quantum number, and each sector has a state of lowest energy, denoted $|S^z\rangle$.

Inserting Eq.~(\ref{SpinOp}) into the Hamiltonian [see Eq.~(1) of the main text], we arrive at the bosonic description of the spin ladder,
\begin{eqnarray}\label{BosonHam}
H &=& \sum_{\alpha=\pm}H(\Phi_\alpha,\Theta_\alpha)=
\int d x \bigg[ \frac{v}{8\pi}\sum_{\alpha=\pm}\big[(\p_x \Theta_\alpha )^2 \cr\cr 
&&\hskip -.05in + ( \p_x \Phi_\alpha )^2 \big] + \sum_{\alpha=\pm} g_\alpha \cos(\Phi_\alpha) + g'   \cos(\Theta_-) \bigg],
\end{eqnarray}
where we drop marginal interactions. The couplings of the non-linear terms are related to the microscopic parameters through $g_\pm \propto (9\Jx/\pi^2 \mp \Jp)$ and $g' \propto 2 \Jp$, and we use symmetric/antisymmetric combinations of the bosonic fields,  $\Phi_\pm = (\Phi_1 \pm \Phi_2)/\sqrt{2}$ and $\Theta_\pm = (\Theta_1 \pm \Theta_2)/\sqrt{2}$. The symmetric sector of Eq.~\fr{BosonHam}, $H_+(\Phi_+,\, \Theta_+)$, is described by an integrable sine-Gordon model. On the other hand, the antisymmetric sector $H_-(\Phi_-,\, \Theta_-)$ is a sine-Gordon model perturbed by an additional operator, the cosine of the dual field.

Having bosonized and changed basis, we proceed to refermionize the theory. This allows us to identify MBSs in the spin chain. To do so, we introduce the right/left ($R/L$) moving fermions (carrying $S^z$ charge)
\be
\psi_{\pm, R} = \frac{\kappa_\pm}{\sqrt{2\pi a_0}} e^{-\frac{i}{2}(\Phi_{\pm}-\Theta_{\pm})}, ~~
\psi_{\pm, L} = \frac{\kappa_\pm}{\sqrt{2\pi a_0}} e^{\frac{i}{2}(\Phi_{\pm}+\Theta_{\pm})}, \nonumber
\ee
where $\kappa_\pm$ are Klein factors that ensure the anti-commutation of fermions of different species, $\{ \kappa_a, \kappa_b\} = \delta_{ab}$. We subsequently express the fermionic fields in terms of their real and imaginary components. With $p=R,L$, we write
\begin{equation}\label{majdef1}
\psi_{+, p} = ( \xi^2_{p} + i\xi^1_p)/{\sqrt2},\quad \psi_{-,p} = (\xi^3_p + i \xi^0_p)/{\sqrt2}.
\end{equation}
We then arrive at a low-energy field theory of Majorana fermions \cite{Shelton} as discussed in the main text.

\subsection{Parity symmetry} 
\label{sec:Parity}

We next explain in more detail how the $\Bbb{Z}_2$ spin parity symmetry discussed in the main text induces a similar $\Bbb{Z}_2$ symmetry in the Majorana system. First consider the smooth part, $M^z_\pm$, of the even and odd combinations of the spin  density operator,
\begin{align}
M^z_{\pm} (x) = \frac{1}{2\pi}\partial_x \Phi_\pm(x) = \frac{1}{2}\sum_{p=L,R} (\psi^\dagger_{\pm, p} \psi^{}_{\pm , p})(x),
\end{align}
expressed both as a fermion density and in terms of the bosonic fields $\Phi_\pm$. Defining quantities integrated
over the system size $L$,
\begin{equation}
\hat S^z_{\pm} =\int^L_0 {\rm d}x M^z_\pm (x),\quad
\hat N_\pm =  \frac{1}{2}\sum_{p}\int_0^L {\rm d}x(\psi^\dagger_{\pm,p} \psi^{}_{\pm, p})(x),
\end{equation}
 and 
$\Delta \Phi_\pm=\Phi_\pm(L)-\Phi_\pm(0)$,
we obtain $\hat S_\pm^z=\hat N_\pm=\frac{1}{2\pi}\Delta \Phi_\pm$. Now consider the global parity constraint, Eq.(2) in the main text, which gives
\begin{equation}
e^{i \pi (\hat S^z_++ \hat S^z_-)}=e^{i\pi (\hat N_+ +\hat N_-)}=1.
\end{equation}
Using Eq.~\eqref{majdef1}, we find
\[
\hat N_+ = -\frac{i}{2}\sum_p \int^L_0 {\rm d}x \xi^1_p\xi^2_p,\quad
\hat N_- = -\frac{i}{2} \sum_p\int^L_0 {\rm d}x \xi^0_p\xi^3_p,
\]
and hence the parity constraint follows in the form
\begin{align}
  e^{\frac{\pi}{2} \sum_{p} \int_0^L {\rm d}x (\xi^1_p\xi^2_p+\xi_p^0\xi_p^3)}=1.
\end{align}

%%%%%%
\section{Ground State Degeneracies  from abelian bosonization}
%%%%%%

In this section, we consider the truncated conformal space approach (TCSA) treatment of the deformed sine-Gordon models in Eq.~\eqref{BosonHam}. Their Hamiltonian density is of the form
\be
{\cal H} =  \frac{v}{8\pi} \bigg[ \Big( \p_x \Theta \Big)^2 + \Big( \p_x \Phi \Big)^2 \bigg]  + g \cos(\Phi) + g' \cos(\Theta) \label{ESG}
\ee
with open boundary conditions.  Our aim is to establish the degeneracies $d_\pm$ listed in Table I of the main text directly from the bosonized field theory.  
As the problem is non-integrable, we require a framework for studying the low-lying states in the spectrum of  Eq.~\fr{ESG}, which is provided by the TCSA.  The TCSA permits a non-perturbative description of perturbed conformal field
theories (such as the sine-Gordon model and its generalizations) \cite{YZ1,*YZ2}.  For a comprehensive review, see Ref.~\onlinecite{review}.  
This approach has been used to study sine-Gordon like models \cite{feverati1998truncated,*feverati1998scaling,*feverati1999non,*bajnok2000k,*bajnok2001nonperturbative,*bajnok2002finite,*bajnok2002spectrum,*bajnok2004susy,*takacs2006double,*toth2004nonperturbative,*palmai2013diagonal,*scnt,*scnt1},  in particular the sine-Gordon model with both Dirichlet~\cite{BajnokNuclPhysB01} and
Neumann boundary conditions~\cite{BajnokNuclPhysB02}. 

We do not provide a full analysis of the phase diagram in Fig.~1 of the main text.  Rather we choose representative points in each phase
to determine the corresponding 
g.s.~degeneracy due to zero modes.  (The latter is not expected to change within a phase since it is tied to signs of fermion masses which are fixed within a phase.)  The points  considered here are 
$(g,g') = (g>0,0),(g<0,0),(0,g'>0),(0,g'<0)$, which correspond to considering the $\cos(\Phi)$ and $\cos(\Theta)$ perturbations
separately.
The TCSA  considers $\cos(\Phi)$ and $\cos(\Theta)$ as perturbations of a free compact boson, using
as a computational basis the Hilbert space of such a boson.  
%%%%%%
\subsection{Bosonic Hilbert Space}
%%%%%%
The Hilbert space for a given bosonic field, $\Phi_\ell$, and its dual, $\Theta_\ell$, on one of the two legs ($\ell=1,2$) of the ladder is understood as follows.  The Hilbert space is divided into sectors marked by their total $S^z$ quantum number.  We denote
the lowest-energy states in such a sector as  
$|S^z\rangle$, with $S^z = 0,\pm 1, \pm 2 \ldots$.
On top of this set of $S^z$-states are states created by acting with oscillator mode operators, $a_{-n}$ (with $n>0$), which appear in the mode expansion of the bosonic fields \cite{EA1992} (we suppress the leg
indices),
\be 
\begin{split}
\Phi(x,t) &= \sqrt{2}\pi + 2^{3/2}\pi \hat S^z \frac{x}{L} \\
& \hskip -.25in + \sum^\infty_{n=1} \frac{2}{n^{1/2}}\sin\Big(\frac{\pi n x}{L}\Big)\Big(a_{n} e^{-\frac{i\pi n t}{L}}+ a_{-n} e^{\frac{i\pi n t}{L}}\Big),\\
\Theta(x,t) &= 2^{3/2}\pi \hat S^z \frac{t}{L} + \Theta_0 \\
& \hskip -.25in + \sum^\infty_{n=1} \frac{2i}{n^{1/2}}\cos\Big(\frac{\pi n x}{L}\Big)\Big(a_{n} e^{-\frac{i\pi n t}{L}} - a_{-n} e^{\frac{i\pi n t}{L}}\Big).
\end{split}
\label{mode_exp}
\ee
The constant term, $\sqrt{2}\pi$, in $\Phi(x,t)$ corresponds to open boundary conditions, where $\Phi$ satisfies Dirichlet boundary conditions. Indeed, putting $\Phi(x=0)=\sqrt{2}\pi$
amounts to identically vanishing lattice spin operators, $S_{\ell,r}^\pm$, at the boundary, see Eq.~\eqref{SpinOp} and Ref.~\onlinecite{EA1992}.
  The zero mode operator $\Theta_0$
appearing in $\Theta(x,t)$ can be considered as the center-of-mass position of the $\Theta$ boson.
(This degree of freedom is absent from the $\Phi$ boson as its boundary conditions have been fixed).
$\Theta_0$ is conjugate to the $\hat S^z$ operator, $[\Theta_0,\hat S^z] = \sqrt{2}i.$
Correspondingly, we see that highest weight sets follow from relations like
$|S^z=\pm 1\rangle = e^{\mp i\Theta_0/\sqrt{2}}|S^z=0\rangle.$
The oscillator modes satisfy the commutation relation
$[a_{n},a_{m}] = n\delta_{nm}$, 
and represent an infinite set of ladder operators. Here, the $a_{-n}$ with $n>0$ are creation operators while the $a_{n}$
annihilate the states $|S^z\rangle$.
The full set of Hilbert space states amounts to products of the creation operators acting on various $|S^z\rangle$ states,
$\prod^N_{i}a_{-n_i}|S^{z}\rangle$ with $n_i > 0.$

\subsection{Truncation and Formation of Hamiltonian Matrix}
The above Hilbert space is infinite dimensional and in practice must be truncated.  Typically this is done by keeping all unperturbed states with energy $E_s$ below some cutoff energy,  $E_s<E_c$.  The unperturbed ($g=g'=0$) energy of a state  $|s\rangle = \prod^{M_s}_{i}a_{-n_i}|S^{z}\rangle$ with  $n_i > 0$  is 
\begin{equation}\label{unperturbed_energies}
E_s = \frac{\pi}{L}\left[\frac{(S^z)^2}{2} + \sum_{i=1}^{M_s} n_i - \frac{1}{24}\right].
\end{equation}
Typically one increases $E_c$ until convergence is obtained (i.e., results become independent of $E_c$),
or until one can detect a trend in the numerical data as a function of $E_c$ so that one can extrapolate (even in principle) $E_c\to\infty$.
There are a variety of ways of performing this extrapolation enhanced by analytical and numerical renormalization group considerations
\cite{review}.  After truncation,
 the Hamiltonian is a finite dimensional matrix whose entries are
determined by the unperturbed energies in Eq.~(\ref{unperturbed_energies}) (on the diagonal) and by matrix elements of the form
$$
\langle s| \cos\big(\Phi(x,0)\big)|s'\rangle, \quad  \langle s| \cos\big(\Theta(x,0)\big)|s'\rangle.
$$
These matrix elements can be easily determined by using the commutators of the oscillator modes with the vertex operators
$e^{\pm i \Phi(x,0)},e^{\pm i \Theta(x,0)}$,
\begin{eqnarray}
[a_n,e^{\pm i \Phi(x,0)}] =  \pm i 2\sin\Big(\frac{\pi n x}{L}\Big)  e^{\pm i \Phi(x,0)},\cr\cr
[a_n,e^{\pm i \Theta(x,0)}] =\pm 2\cos\Big(\frac{\pi n x}{L}\Big)  e^{\pm i \Theta (x,0)} , \nonumber
\end{eqnarray}
together with the fundamental matrix elements of the vertex operators on the highest weight $S^z$ states,
\begin{eqnarray}
\langle S^z|e^{\pm i\Phi(x,0)}|S^{z\prime}\rangle &=& \delta_{S^z,S^{z\prime}},\cr\cr
\langle S^z|e^{\pm i\Theta(x,0)}|S^{z\prime}\rangle &=& \delta_{S^z,S^{z\prime}\mp 2}. \nonumber
\end{eqnarray}
Once the Hamiltonian matrix has been computed, it can be easily numerically diagonalized and the resulting spectrum extracted.

For studying the $\cos(\Phi)$ perturbation we will pursue the simple strategy of forming the computational basis by
truncating the unperturbed spectrum for different values
of $E_c$ and seeing whether we see g.s. degeneracies develop (or not) as $E_c$ is increased.  However for the $\cos(\Theta)$ study,
we will alter the strategy somewhat.  We have found that keeping a large, fixed number of highest weight states, 
$\big\{|-S^z_\text{max}\rangle, |-S^z_\text{max}+1\rangle,\cdots,|S^z_\text{max}\rangle\big\}$ while
truncating at different levels the oscillator content works best.  This then involves keeping states of the form
\be
|s\rangle = \prod^{M_s}_{i}a_{-n_i}|S^z\rangle;\qquad
\sum_{i=1}^{M_s} n_i \leq N, \quad |S^z| \leq S^z_\text{max},
\nonumber
\ee
for different choices of $N$ and $S^z_\text{max}$.
It is similar to a truncation of states in terms of energy, but we do not count the contribution of finite $S^z$ to a state's energy.
This strategy works here as the $\cos(\Theta)$ perturbation connects states with different values of $S^z$, and the physics is dominated by the zero mode $\Theta_0$. The problem thus effectively becomes 0+1 dimensional, where
the oscillator modes only renormalize
the underlying zero-mode problem in a quantitative (not qualitative) fashion.

\subsection{Analysis of the $\cos(\Phi)$ Perturbation}

\begin{figure}
\includegraphics[width=0.37\textwidth,trim = {5 0 0 24},clip]{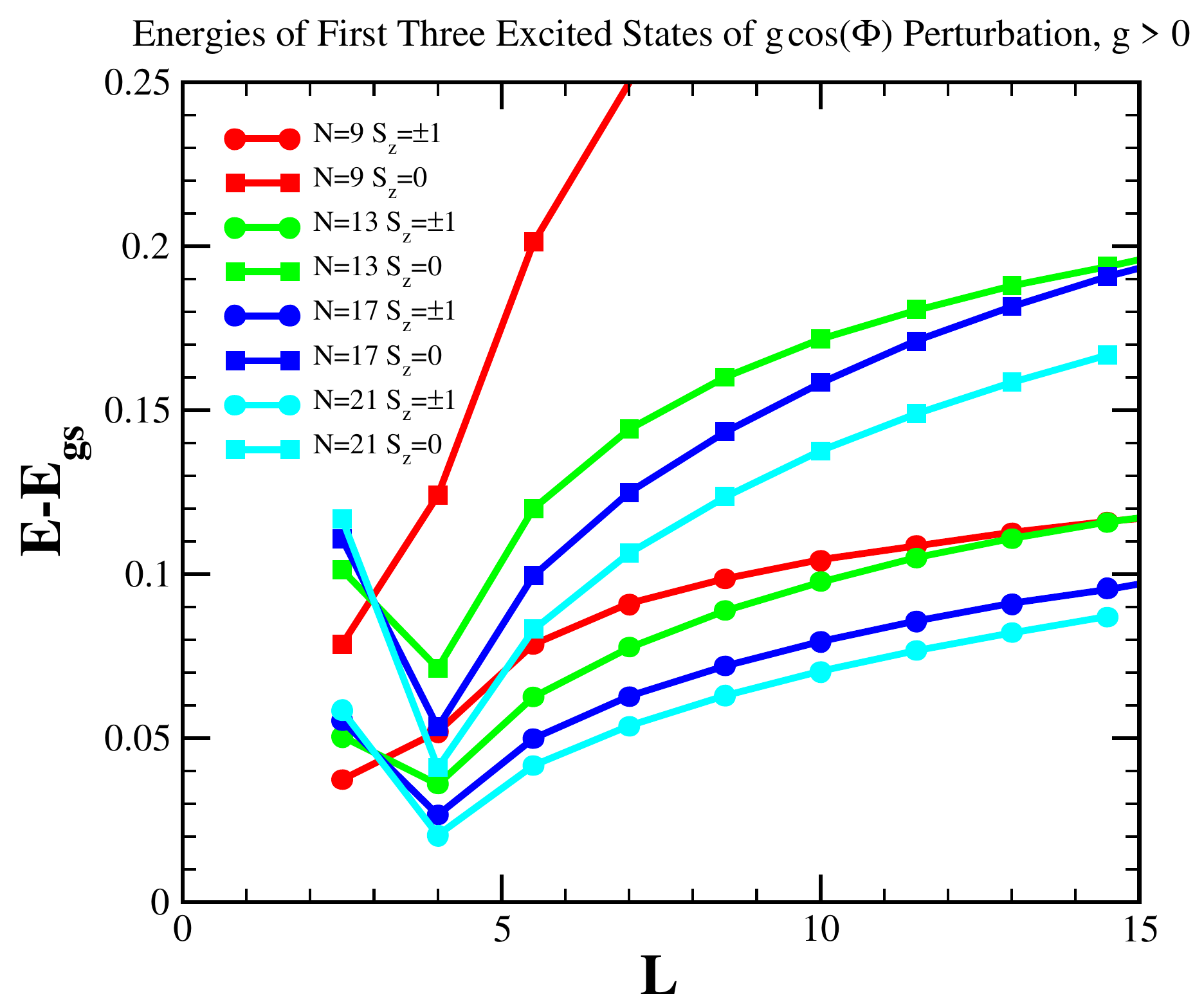}
\caption{TCSA data for the energies of the three lowest-lying states
for a pure $g\cos (\Phi)$ perturbation with $g > 0$.  We expect these three states
to be degenerate for $L\to \infty$ and without truncation ($N\to \infty$).  The first two of these states are degenerate carrying $S^z=\pm1$. The third carries $S^z=0$ and is the highest energy of the
three for finite $L$ and finite $N$.
We present data for four different cutoffs ($N=9,13,17,$ and $21$) and system sizes $L$
(ranging from $2.5$ to $14.5$). We see that as $N$ increases, the gap between these states
and the g.s.~decreases over a range of $L$, especially for large $L$.}
\label{Fig:TCSA_data_phi_pos}
\end{figure}

We now will consider the Hamiltonian, see Eq.~\eqref{ESG}, for $g'=0$, where we have a pure $\cos(\Phi)$ perturbation.
For $g>0$, we expect the g.s.~to have a 4-fold degeneracy
while for $g<0$, the g.s.~should be unique.  In Tab.~I of the main text, this covers all four instances of the even sector
($d_+=4$ or $d_+=1$) and two instances of the odd sector (for the \VBSp and \VBSm phases). 
In Fig.~\ref{Fig:TCSA_data_phi_pos} we present our numerical data 
for the energies of the three lowest excited states, for $g>0$.  Here $g$ has been chosen so that the bulk gap equals unity.  The excited states are labelled by the $S^z$ quantum number
of the sector in which they lie.  The first two states are found in the $S^z=\pm 1$ sectors and are degenerate, while the third excited
state is in the $S^z = 0$ sector.  We present data for a number of different energy truncations as marked by $N$, related to $E_c$ via
$E_c = (\pi/L)\left[(S^z)^2 /2 + N - 1/24 \right].$
We plot this data vs the chain length $L$.  At small $L$, we are in the UV limit and expect energy levels $\sim 1/L$.  While we do not present data for very small $L$, this trend is observable around $L\sim 4$ for large $N$. In an intermediate range,  $L\approx 4$ to $6$, we expect the
low-lying states to have roughly the same energies as for $L\to \infty$.  At larger values of $L$,
we expect the appearance of finite truncation effects which manifest themselves as increases in the energies of the lowest
lying states relative to
the g.s.~energy.  We see this in Fig.~\ref{Fig:TCSA_data_phi_pos} for $L>6$.  Of course as $N$ is increased,
we expect the data at larger $L$ to tend to return towards the values obtained in the intermediate $L$ region.  And this trend we 
indeed do see in the data as well.
Overall the data presented in Fig.~\ref{Fig:TCSA_data_phi_pos} allows us to conclude that the system develops a 4-fold degenerate
g.s.~as asserted.  We can clearly see that in the intermediate region ($L=4$ to $6$), as $N$ is increased, both the first excited state in the $S^z=0$ sector as well as the lowest lying states in $S^z=\pm 1$ sectors  become degenerate with the $S^z=0$ state.

\begin{figure}
\includegraphics[width=0.35\textwidth,trim={5 0 0 24},clip]{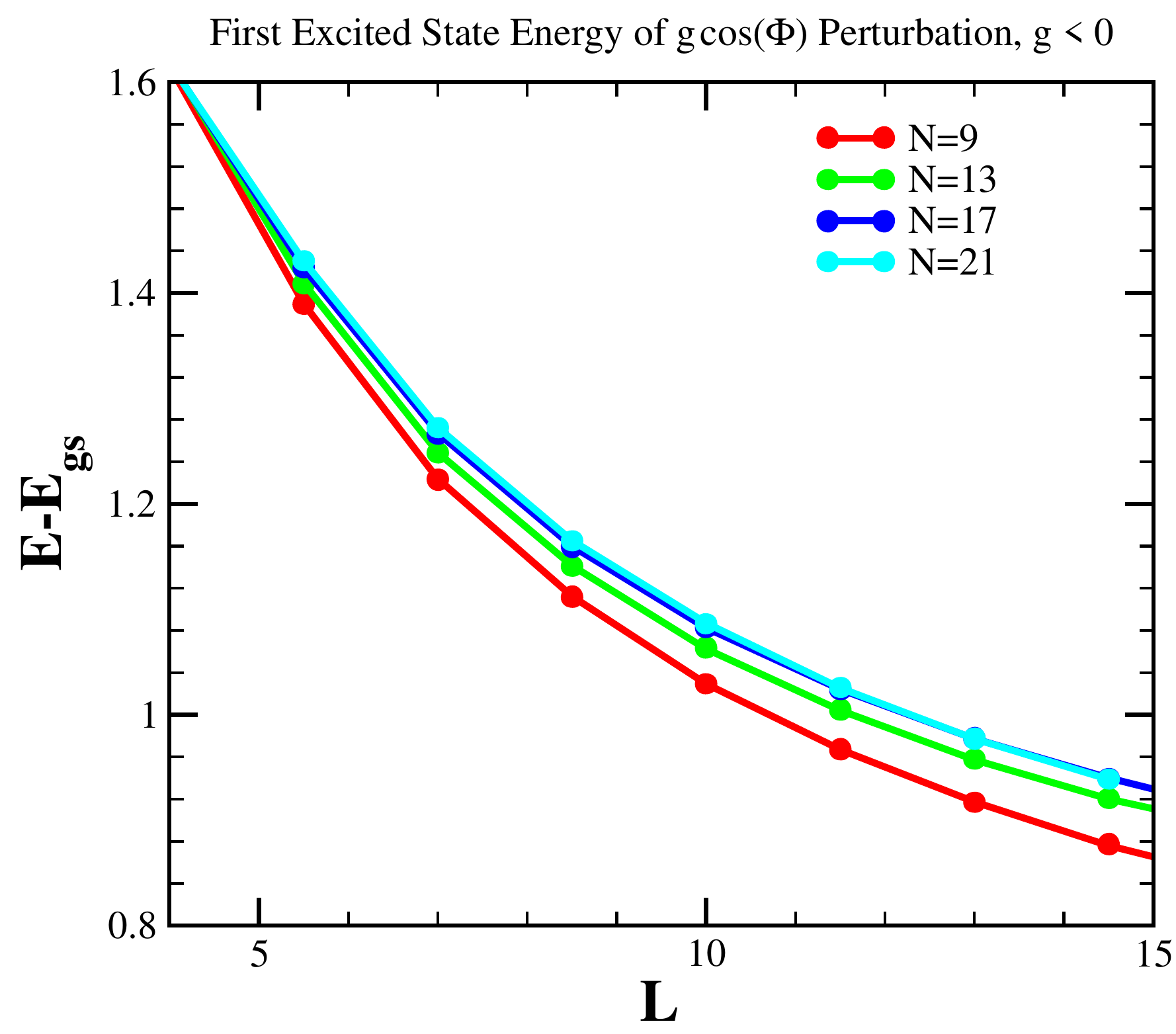}
\caption{TCSA data for the energy of the first two (degenerate) excited states for a pure
$g\cos (\Phi)$ perturbation with  $g< 0$.  These excited states correspond to $S^z=\pm 1$
solitons along our system. 
We present data for four different cutoffs ($N=9,13,17$, and $21$), with system size $L$ ranging from $2.5$ to $14.5$.
We see with increasing $N$, the energy approaches the bulk gap 
value ($E=1$) over an increasingly wide range of $L$.}
\label{Fig:TCSA_data_phi_neg}
\end{figure}

In Fig.~\ref{Fig:TCSA_data_phi_neg} we present our TCSA data for $g<0$.  Here we have again have chosen the value of $g$ so that the gap in the bulk is unity.  And because there should be no g.s.~degeneracies in this case, we expect the energies of the first two excited states here to
be degenerate and equal to 1.  And this is what we see.  In comparison to the $g>0$ case, the region of $L$ where the conformal ($g=0$) UV physics dominates is now larger, extending to $L \sim 6$.  But for $L>6$, the energy of the first excited states approaches $1$.  
As we go to larger $L$ and see the effects of finite truncation, the energy of degenerate excited states
dips below 1.  But as the cutoff $N$ is increased, the energy returns to $1$, albeit slowly.  This data is then consistent with a unique g.s.~for $g<0$.

%%%%
\subsection{Analysis of the $\cos(\Theta)$ Perturbation}
%%%%

\begin{figure}
\includegraphics[width=0.35\textwidth, trim = {5 0 0 26}, clip]{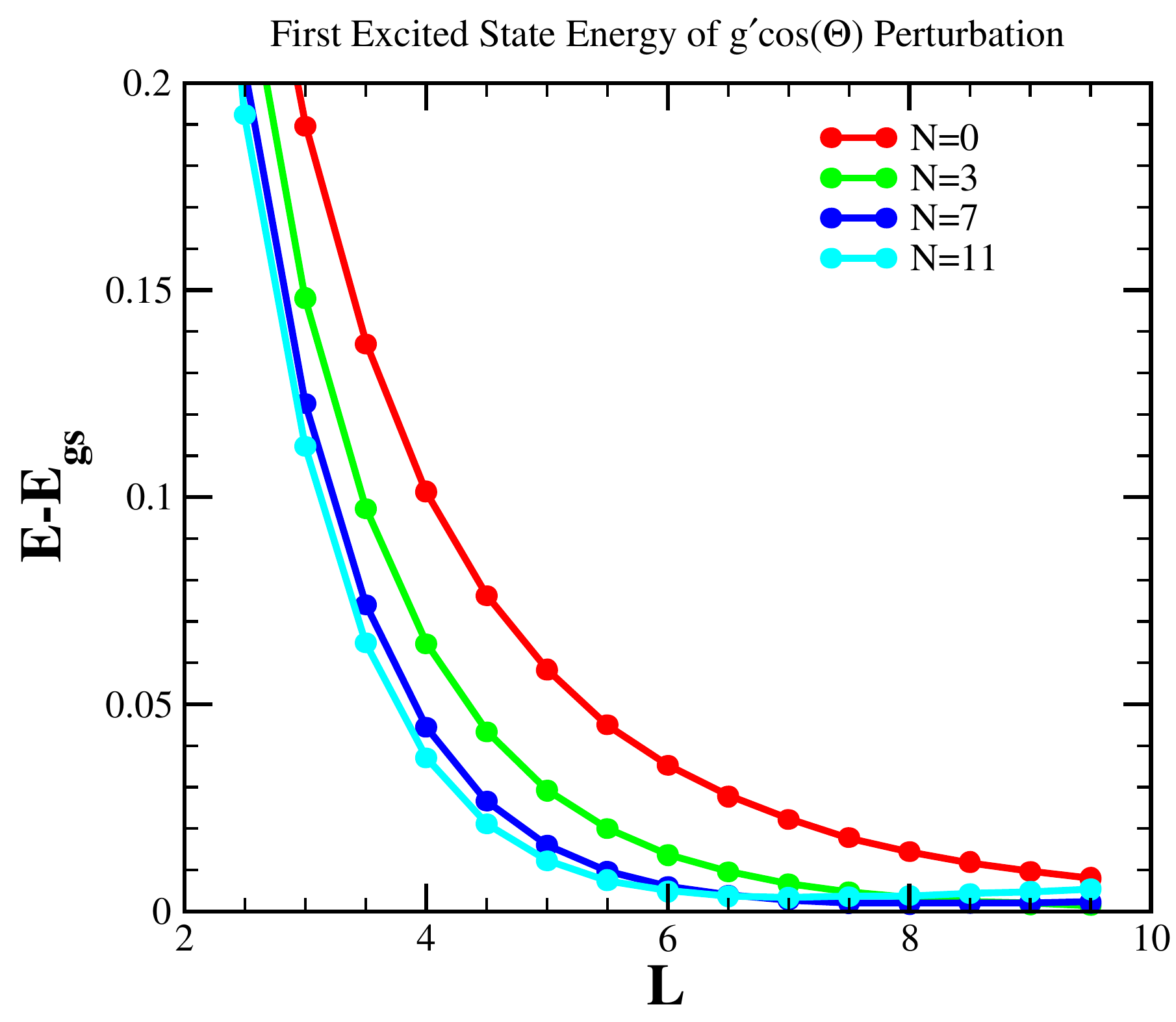}
\caption{TCSA data for the energy of the first excited state for a pure $g'\cos(\Theta_-)$
perturbation with $g'>0$.
We plot the data for a fixed $S^z_\text{max}=6$ (see text)
while varying the oscillator mode content of the truncated Hilbert space from no oscillator modes ($N=0$) to
$N=11$.  We see that the first excited state becomes degenerate with the g.s.~for $L>4$ as $N$ is increased.}
\label{Fig:TCSA_data_theta_pos}
\end{figure}

We now turn to the consideration where the theory is
perturbed purely by the dual boson, $g'\cos(\Theta)$, in Eq.~\eqref{ESG}.
Unlike with the $\cos(\Phi)$ perturbation,
the g.s.~degeneracy does not depend on the sign of $g'$, and we thus only consider the case $g'>0$.
Again, we choose $g'$ such that the bulk gap is 1.
We expect a 2-fold g.s.~degeneracy which corresponds to the $-$ sector for the H and RS phases.
We present our data in Fig.~\ref{Fig:TCSA_data_theta_pos}. For $L>6$, we exit the UV regime
where conformal physics dominates and the gap to the first excited state vanishes. The region in $L$ over which
the gap vanishes increases as the cutoff $N$ increases.
We here have used a modified cutoff strategy where we leave the number of highest weight states $|S^z\rangle$ fixed with $S^z_\text{max}=6$,
regardless of the value of $N$.  
We then vary $N$ and allow the oscillator content of the states built on top of this set of $|S^z\rangle$-states
to change.   We see from Fig.~\ref{Fig:TCSA_data_phi_neg} that even if we consider a truncation of the Hilbert space that is pure
highest weight states (i.e. $N=0$), the results are not terrible -- we find a gap below $0.05$ in our units.  As we then allow for $N>0$, this already very small gap rapidly decreases to zero.

\subsection{Bosonic Phase Diagram}\label{sec2e}

\begin{figure}
\includegraphics[width=0.45\textwidth,trim={0 0 0 0},clip]{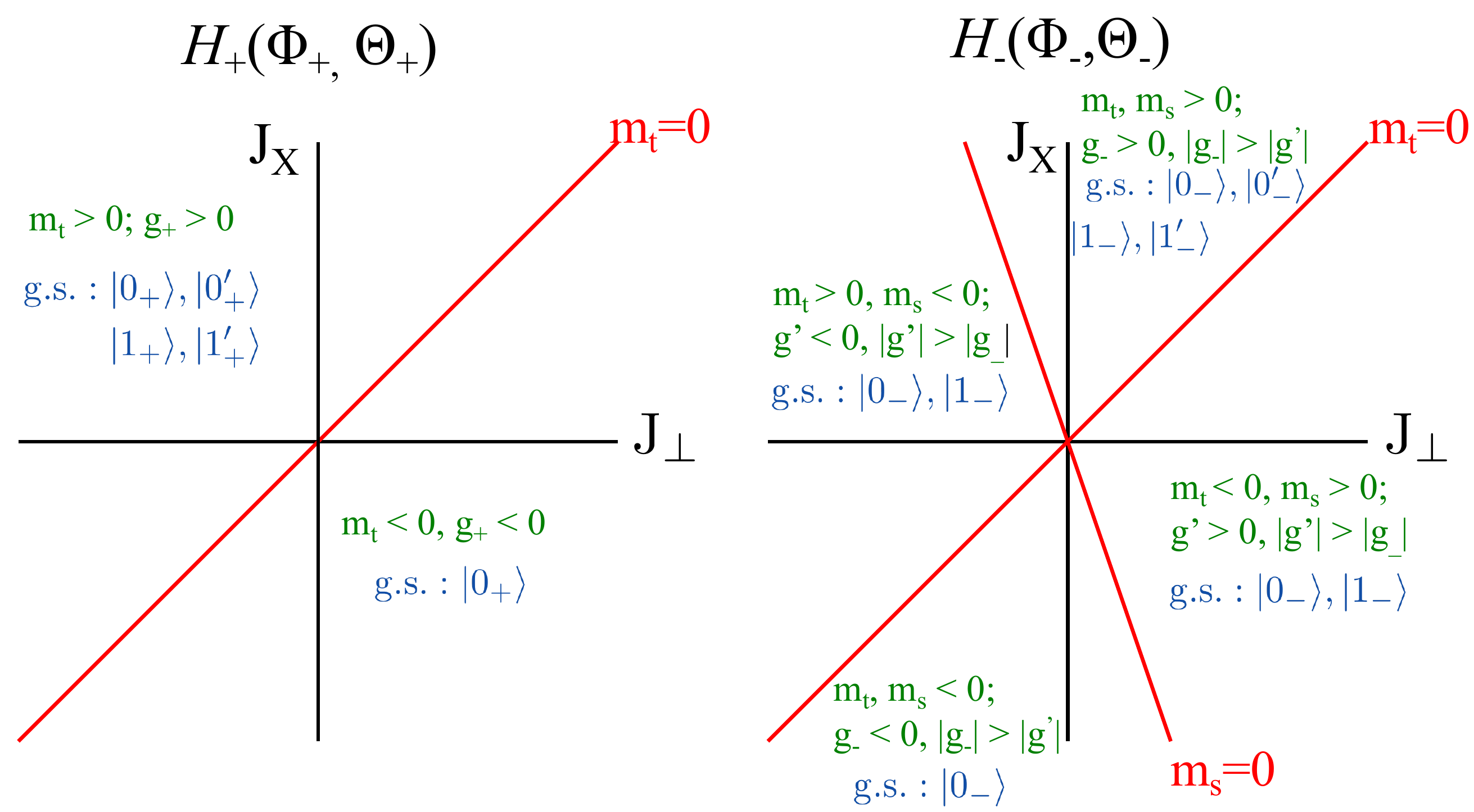}
\caption{Phase diagrams for the two bosonic sectors, $H_\alpha(\Phi_\alpha,\Theta_\alpha)$, $\alpha=\pm$.}
\label{Fig:BosonPhaseDiag}
\end{figure}

In Fig.~\ref{Fig:BosonPhaseDiag} we summarize the results of our TCSA analysis.  We show both the g.s.~degeneracies for the bosonic Hamiltonians of the even and odd sectors.  In this diagram we have labelled the degenerate ground states according to their parity.  So, for example, for the even 
sector Hamiltonian $H_+$ with $g_+>0$, there are four degenerate
ground states, two with even parity, $|0_+\rangle,|0_+'\rangle$, and two with odd parity, $|1_+\rangle,|1_+'\rangle$.

Let us now consider how taking into account parity restricts the g.s. manifold of the full ladder (which comes from tensoring ground states of the even and odd sectors together).  Take the \VBSp phase, where $m_t,m_s > 0$. 
For the $+$~sector, the bosonic g.s.~is 4-fold degenerate.  Similarly, the g.s.~in the $-$~sector is also 4-fold degenerate, with two states of each parity: $|0_-\rangle,|0'_-\rangle,|1_-\rangle,|1'_-\rangle$. 
The gluing rules matching parity then permit the following g.s.~manifold for the \VBSp phase:
\be
\begin{split}
&|0_+;0_-\rangle,~|0_+;0'_-\rangle,~|0'_+;0_-\rangle,~|0'_+;0'_-\rangle,\nn
&|1_+;1_-\rangle,~|1_+;1'_-\rangle,~|1'_+;1_-\rangle,~|1'_+;1'_-\rangle. 
\end{split}
\ee
States such as $|0_+;1_-\rangle$ are disallowed because the $+$ and $-$ sectors have different parities and hence the \VBSp phase has an 8-fold (not 16-fold) degenerate g.s.~in agreement with DMRG.

As a second example, the H phase has $m_t>0,\,m_s<0$. 
The $+$~sector has the g.s. manifold $|0_+\rangle,|0'_+\rangle,|1_+\rangle,|1'_+\rangle$, while in the $-$~sector we have only $|0_-\rangle,|1_-\rangle$.
Thus the permitted g.s.~set is given by
\begin{equation}
|0_+;0_-\rangle,\quad|0'_+;0_-\rangle,\quad|1_+;1_-\rangle,\quad|1'_+;1_-\rangle \nonumber
\end{equation}
which is 4-fold degenerate, consistent with DMRG.

%%%%%%%%%%%%%%%%%%%%%%%%%%%%%%%%%5%%%%%%%%%
\section{Splitting of Ground State  Degeneracy for Sharp Transitions}
%%%%%%

Using the notation in Sec.~\ref{sec2e}, we 
 the two ground states of the RS--VBS$_-$--RS ladder are given by 
 \begin{eqnarray}
|{\rm gs}1\rangle &\equiv& |0_+,0_+,0_+;0_-,0_-,0_-\rangle,\cr\cr
|{\rm gs}2\rangle &\equiv& |0_+,0_+,0_+;1_-,0_-,1_-\rangle. \nonumber
\end{eqnarray}
Adding the g.s.~parities 
of each individual portion of the ladder (modulo 2), the $\pm$ sectors have equal (and 
even) parity.
Now it is clear in the lattice spin picture why the degeneracy of the two ground states in the RS--VBS$_-$--RS ladder
is broken.  As shown in Fig.~2 of the main text, it is only with soft boundary conditions that the exact position of the 
singlets of the VBS$_-$ phase along the length of the ladder is ambiguous (up to a single lattice spacing), thus leading to a two-fold degeneracy.
However it is also useful to understand why the soft boundary conditions are needed for degeneracy in the Majorana fermion languagre.
Nominally, the sharpness of the boundary is a local perturbation which is not expected not break the degeneracy of states involving spatially separated MBSs.  The key to resolving this conundrum is that a perturbation that is local in spin
operators will not necessarily be local in the Majorana language.

\begin{figure}[t]
\includegraphics[width=0.45\textwidth]{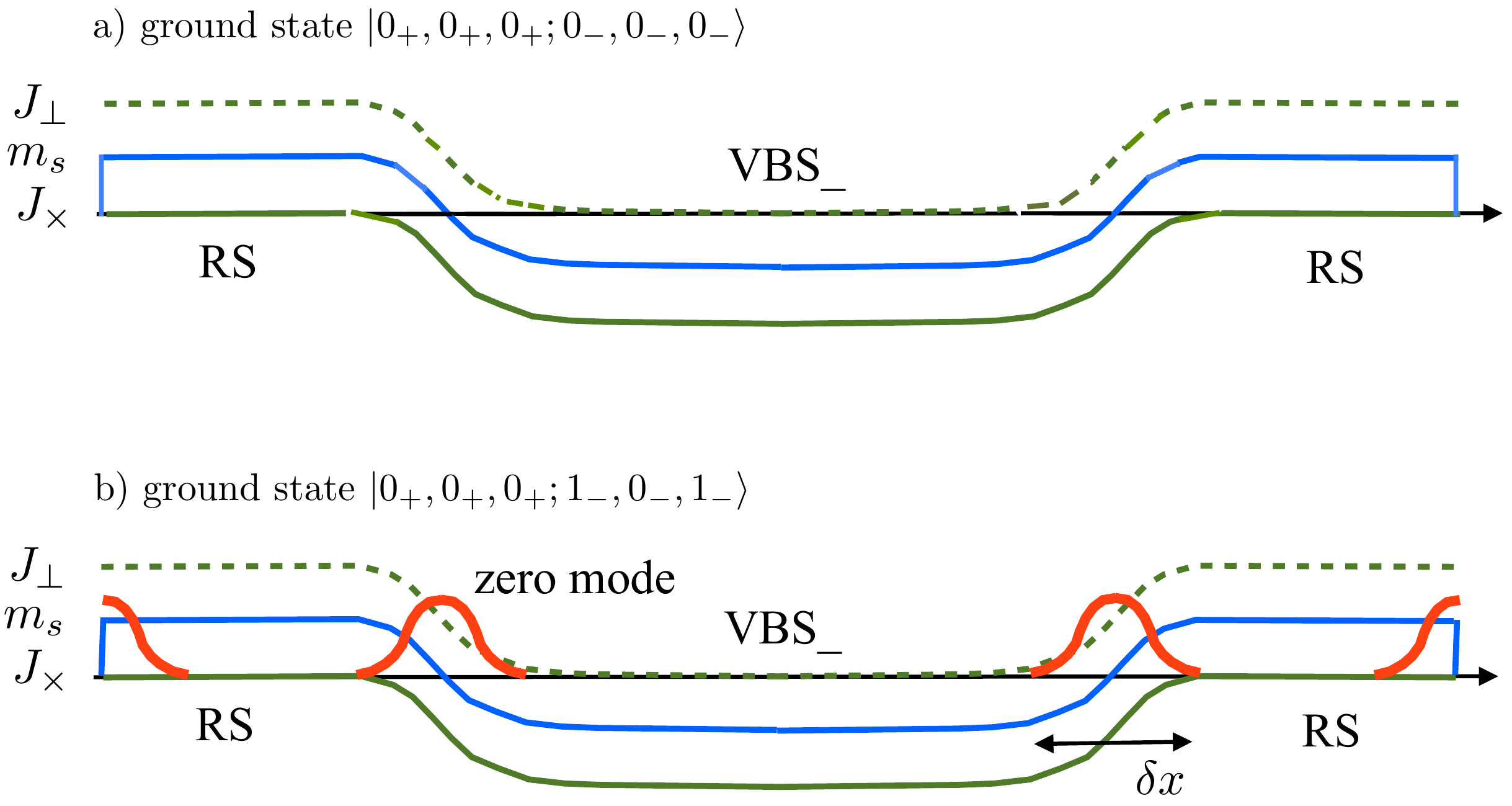}
\caption{The two possible degenerate ground states for the
  RS--VBS$_-$--RS ladder.  The first,
  $|0_+,0_+,0_+;0_-,0_-,0_-\rangle$, has no MBSs while the second, $|0_+,0_+,0_+;1_-,0_-,1_-\rangle$, has four MBSs.}
\label{Fig:2fold_gs}
\end{figure}

To be clear, Fig.~\ref{Fig:2fold_gs} shows the possible configuration of the zero modes along the ladder for the two possible ground states of the RS--VBS$_-$--RS ladder.  The $|{\rm gs}1\rangle$ state has no zero modes
present, while the $|{\rm gs}2\rangle$ state has four zero modes: one localized at each end of the ladder, and one at each of the RS--VBS$_-$ interfaces. The parity selection rule here amounts to forbidding states with only two zero modes.

To see how sharp variations in the spin ladder parameters can induce a non-local perturbation in terms of the Majorana fermions, we first
need to write all of the spin operators in bosonized/fermionic form.  Each spin ${\bm S_{\ell,r}}$ has a smooth $k=0$ part, ${\bm M}_{\ell,r}$, and a staggered
$k=\pi$ part, ${\bm N}_{\ell,r}$. 
The even and odd combinations of  spin operators across a given rung,
\begin{equation}
{\bm M}_{\pm,r} = {\bm M}_{1,r} \pm {\bm M}_{2,r}, \quad
{\bm N}_{\pm,r} = {\bm N}_{1,r} \pm {\bm N}_{2,r},
\end{equation}
can be written in terms of the operators describing 
the four copies of the quantum Ising model forming the field theoretic representation
of the spin ladder. With $a=1,2,3$, we have
\begin{eqnarray}
M^{a}_{+,r} &\sim& \epsilon^{abc}\Big(\xi_R^b(x_r)\xi_R^c(x_r) + \xi_L^b(x_r)\xi_L^c(x_r)\Big),\cr\cr
M^{a}_{-,r} &\sim& \xi_R^0(x_r)\xi_R^a(x_r) + \xi_L^0(x_r)\xi_L^a(x_r),\cr\cr
N^1_{+,r} &\sim& \cos\left(\frac{\Theta_+(x_r)}{2}\right)\cos\left(\frac{\Theta_-(x_r)}{2}\right) \cr\cr 
  &\sim& \mu^0(x_r)\mu^1(x_r)\sigma^2(x_r)\sigma^3(x_r),\cr\cr
N^2_{+,r} &\sim& \sin\left(\frac{\Theta_+(x_r)}{2}\right)\cos\left(\frac{\Theta_-(x_r)}{2}\right) \cr\cr 
  &\sim&\mu^0(x_r)\sigma^1(x_r)\mu^2(x_r)\sigma^3(x_r),\cr\cr
N^3_{+,r} &\sim& \sin \left(\frac{\Phi_+(x_r)}{2}\right)\cos\left(\frac{\Phi_-(x_r)}{2}\right) \cr\cr 
  &\sim& \mu^0(x_r)\sigma^1(x_r)\sigma^2(x_r)\mu^3(x_r),\cr\cr
N^1_{-,r} &\sim& \sin\left(\frac{\Theta_+(x_r)}{2} \right)\sin\left(\frac{\Theta_-(x_r)}{2}\right) \cr\cr
  &\sim&\sigma^0(x_r)\sigma^1(x_r)\mu^2(x_r)\mu^3(x_r),\cr\cr
N^2_{-.r} &\sim& \cos\left(\frac{\Theta_+(x_r)}{2}\right)\sin\left(\frac{\Theta_-(x_r)}{2}\right) \cr\cr 
  &\sim& \sigma^0(x_r)\mu^1(x_r)\sigma^2(x_r)\mu^3(x_r),\cr\cr
N^3_{-,r} &\sim& \cos\left(\frac{\Phi_+(x_r)}{2}\right)\sin\left(\frac{\Phi_-(x_r)}{2}\right) \cr\cr 
  &\sim& \sigma^0(x_r)\mu^1(x_r)\mu^2(x_r)\sigma^3(x_r).
\end{eqnarray}
The fermionic fields $\xi^{b}_{L,R}$ ($b=0,1,2,3$) are introduced in the main text. For each of the four copies of the (fermionic)
Ising theories,  we have associated spin (order) and disorder fields, $\sigma^{b}$ and $\mu^{b}$, respectively. 
It is crucial here that the operators $\sigma^b$ and $\mu^b$ are \textit{non-local in terms of the fermions $\xi^b_{L,R}$}.  

If the (fermionic) Ising theory is in its ordered phase ($m>0$), there will be non-zero matrix elements of the spin field in the ground state manifold while the disorder operator in this same manifold vanishes,
\be
\la \text{gs\#} | \s^j | \text{gs\#}\ra \Big\vert_{m_j > 0} \neq 0, \quad \la \text{gs\#} | \mu^j | \text{gs\#}\ra  \Big\vert_{m_j > 0} = 0. \nonumber
\ee
If instead the theory is in its disordered phase, $m < 0$, the situation is reversed: matrix elements of the 
disorder operator can be non-zero while those of the spin operator are identically zero,
\be
\la \text{gs\#} | \s^j | \text{gs\#}\ra \Big\vert_{m_j < 0} = 0, \quad \la \text{gs\#} | \mu^j | \text{gs\#}\ra  \Big\vert_{m_j < 0} \neq 0. \nonumber
\ee
Let us now consider how these matrix elements may cause a splitting of the g.s.~degeneracy.
In a ladder that is either translationally invariant or has smooth variations (whose length scale is far greater
than the lattice spacing), the smooth ($\bm{M}$) and staggered ($\bm N$) parts of the spin operators do not couple in the Hamiltonian. Indeed, such terms rapidly oscillate and average to zero under the spatial integral. However, if the exchange couplings vary on the order of the lattice spacing, terms such as 
${\bm M}_+\cdot {\bm N_+}$ and/or  ${\bm M}_-\cdot {\bm N_-}$
can appear in the Hamiltonian. 
Using the operator product expansion
$\sigma \cdot \xi_{L,R} \sim \mu$,
we see that the following terms can then appear in the low-energy effective theory:
\begin{eqnarray}
{\bm M}_+\cdot {\bm N_+} \sim  {\bm M}_-\cdot {\bm N_-} &\sim& \cos(\Phi_+/2)\cos(\Phi_-/2)\cr\cr
&\sim& \mu^0\mu^1\mu^2\mu^3.
\end{eqnarray}
Both of these lattice terms ($\pm$) have the same operator form in the continuum. 

Now how does ${\bm M_\pm}\cdot{\bm N_\pm}$ lead to a splitting of the putative g.s.~degeneracy 
between $|{\rm gs1}\rangle$ and $|{\rm gs 2}\rangle$? The easiest way to see this is to notice that the
singlet patterns of the states $|{\rm gs1}\rangle$ and $|{\rm gs2}\rangle$ in the VBS$_-$ portion of the ladder are shifted by one lattice spacing relative to one another.  
Moreover, under a shift by one lattice spacing, the bosonic fields are correspondingly shifted as
$\Phi_+ \rightarrow \Phi_+ + 2\pi$ and $\Theta_+ \rightarrow \Theta_+ + 2\pi$ while 
$\Phi_- \rightarrow \Phi_-$ 
and $\Theta_- \rightarrow \Theta_-$.
Using the bosonic form of ${\bm M_\pm}\cdot{\bm N_\pm}$, this implies
\begin{eqnarray}
\langle {\rm gs1}|{\bm M_\pm}\cdot{\bm N_\pm}(x)|{\rm gs1}\rangle = -\langle {\rm gs2}|{\bm M_\pm}\cdot{\bm N_\pm}(x)|{\rm gs2}\rangle ,\cr
\end{eqnarray}
where $x$ is in the VBS$_-$ segment of the inhomogeneous ladder.
As ${\bm M_\pm}\cdot{\bm N_\pm} \sim \mu_0\mu_1\mu_2\mu_3$,  these matrix elements are non-zero since in the VBS$_-$ phase, all fermion masses are negative.

The RS--VBS$_-$--RS ladder with rapidly varying couplings therefore has additional Hamiltonian terms  of the form
\begin{eqnarray}
\delta H &=& \alpha\int_{{\rm left~interface}} \hskip -.5in d x\, \mu^0(x)\mu^1(x)\mu^2(x)\mu^3(x) \cr\cr 
&&\hskip -.1in + \alpha \int_{{\rm right~interface}} \hskip -.5in  d x\, \mu^0(x)\mu^1(x)\mu^2(x)\mu^3(x) ,
\end{eqnarray}
where the spatial integrals are confined to the boundary regions between the phases, averaging to zero otherwise. From the above discussion we expect that $\delta H$ induces a splitting in energy of the two `ground states' proportional to the coupling $\alpha$ at first
order in perturbation theory.
Thus sharp boundaries between phases in the spin model lead to a splitting of the degeneracy in the Majorana theory.

\begin{figure}
\includegraphics[width=\linewidth]{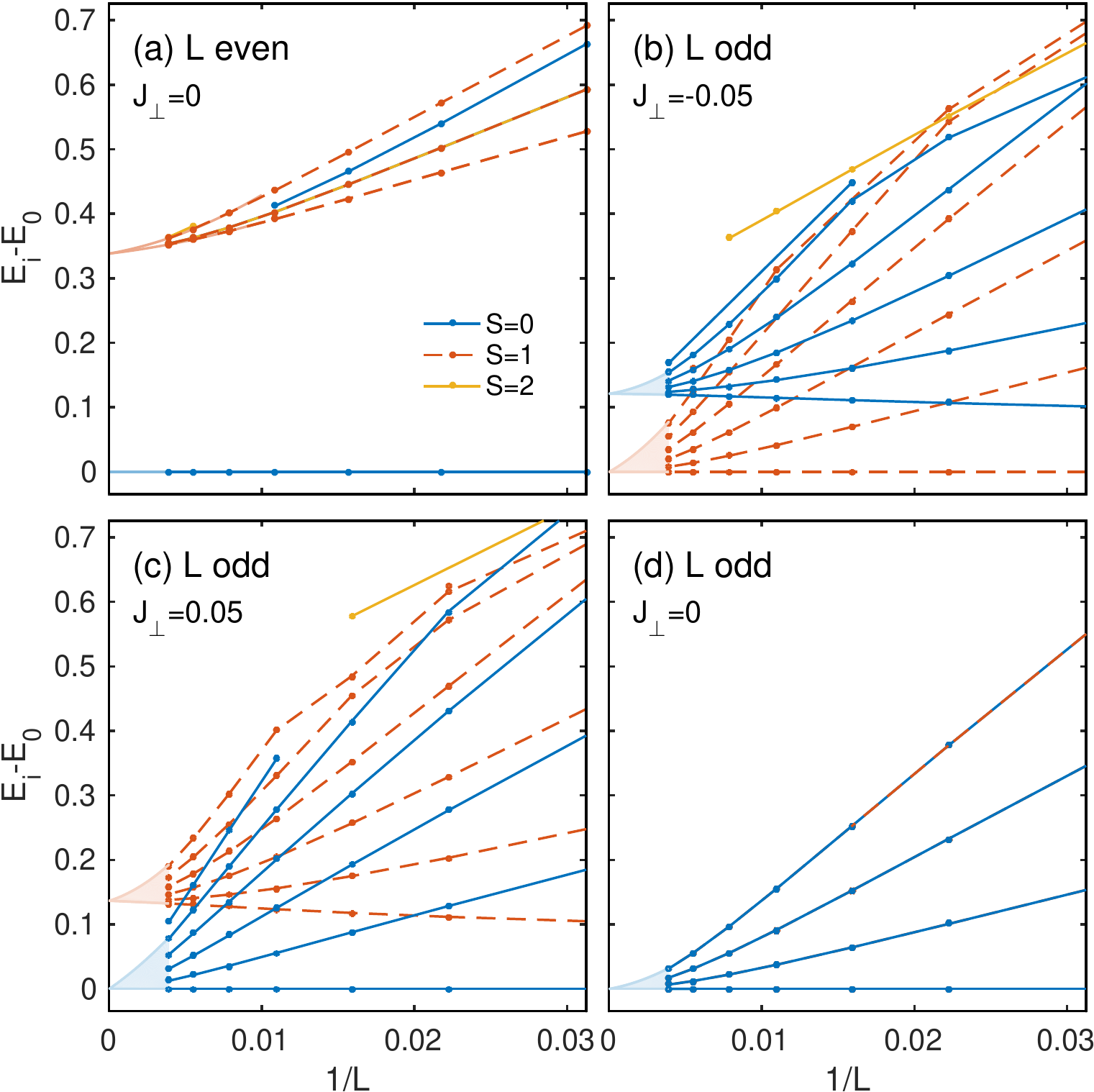}
\caption{Finite size scaling
  $1/L\to 0$ of the DMRG low-energy eigenstates in the uniform
  \VBSm ladder close to $\Jp=0$ for ladders of even
  [panel (a)] and odd lengths [panels (b-d)].
  At least $N_\psi \geq 8$ low-energy
  symmetry multiplets are targeted. Lines of the 
  same color belong to the same global symmetry 
  sector as indicated in the legend of (a).
}
\label{Fig:VBSm}
\end{figure}

\begin{figure}
\includegraphics[width=\linewidth]{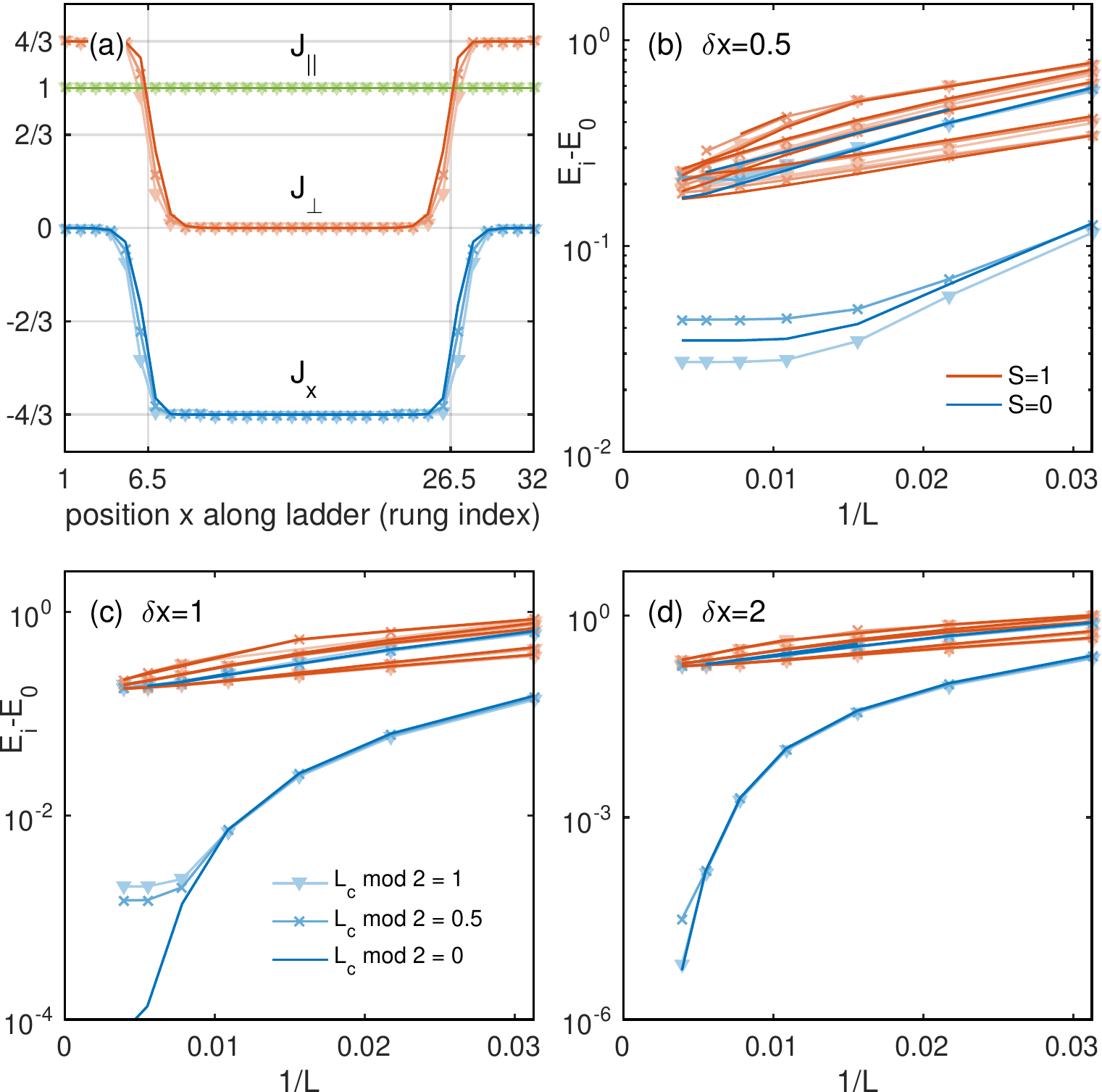}
\caption{ Finite size scaling of low-energy
  DMRG eigenstates in the RS--\VBSm--RS ladder.
  (a) The couplings are varied as in 
  $\Jp(x) = \tfrac{4}{3} (1-w(x))$ and $\Jx(x) =-\tfrac{4}{3} w(x)$,
  with $w(x)$  in \Eq{eq:window:wx}.
  Panels (b-d) show the finite size analysis $1/L \to 0$
  for $\delta x=0.5$, $1$, and $2$, respectively,
  with $L=32 \ldots 256$.
  The colors specify the symmetry sectors
  [singlet sector $S=0$ in blue, and triplet sector $S=1$
  in red, cf.~legend to (b)].
  Each panel contains data
  from three slightly different systems to
  explore (the eventually minor)
  even-odd effects in the length of the \VBSm
  center region [see legend to (c)].
}
\label{Fig:RS-VBSm-RS-ext}
\end{figure}

\begin{figure}
\includegraphics[width=1\linewidth]{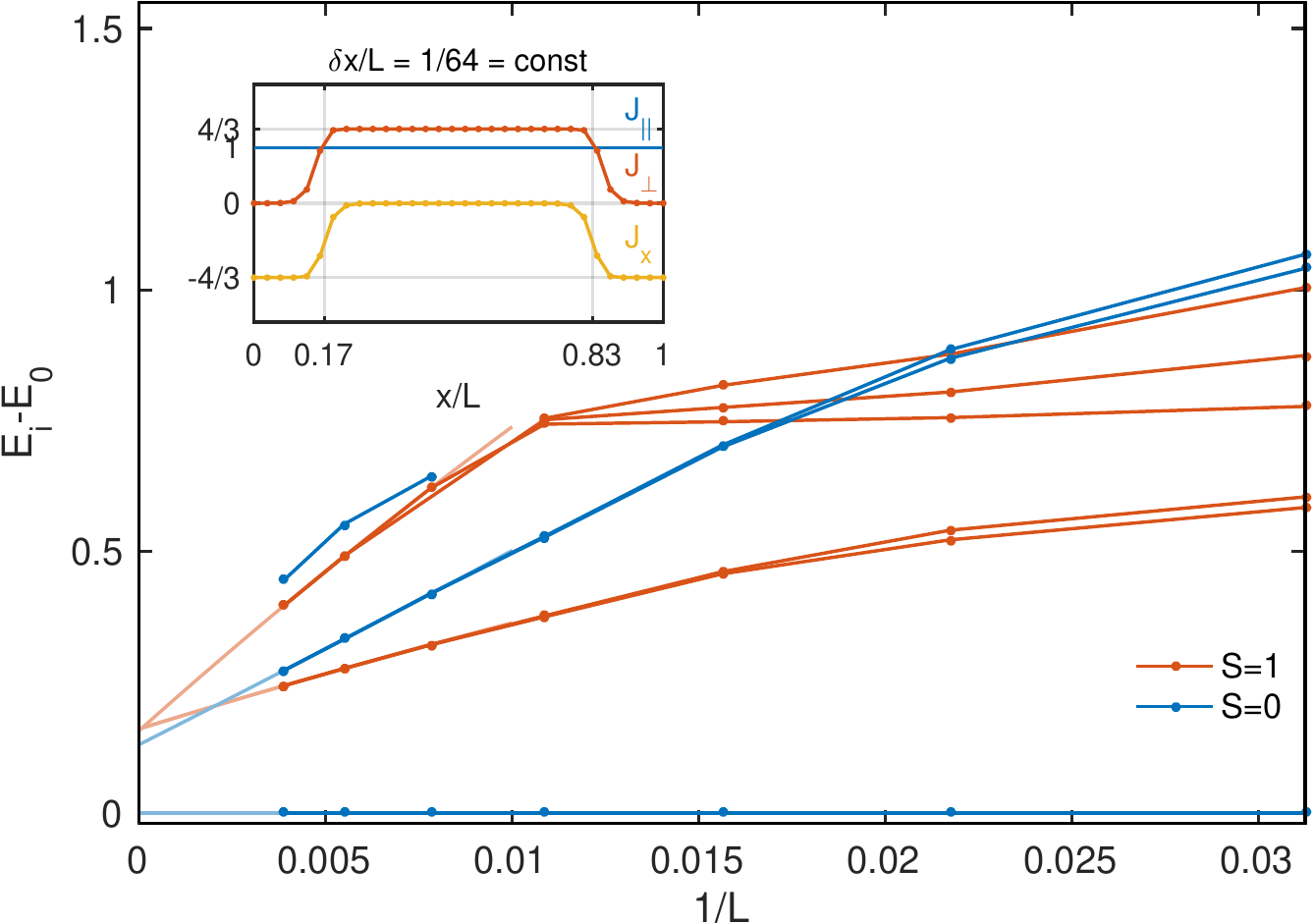}
\caption{Finite-size analysis of the low-energy
   DMRG eigenstates for the \VBSm--RS--\VBSm ladder
   plotted vs. $1/L$ for
   $L=32,\ldots,256$.
   We used $\delta x = L/64 =\mathrm{const}$ and
   $(\Jp,\Jx)=\tfrac{4}{3}(\sin \varphi, -\cos\varphi)$
   where $\varphi = \pi w/2$
   and $w(x)$ as in \Eq{eq:transition:fx}.
   The parameter profile is shown
   in the inset for $L=32$.
   Global symmetry sectors
   are again indicated by color (see legend),
   with the data tentatively extrapolated to
   $1/L\to0$ (lines in light colors).
   As $L$ increases, the transition becomes smoother.
   Yet for all system sizes analyzed
   for a transition width up to $\delta x=2$,
   the g.s. clearly remains a unique singlet.
}
\label{Fig:DMRG_VBSmRS}
\end{figure}

%%%%%%%%%%%%%%%%%%%%%%%%%%%
\section{DMRG Background and Further Results
\label{app:DMRG}}
%%%%%%%%%%%%%%%%%%%%%%%%%%%

The DMRG~\cite{White92,*Schollwoeck05,*Schollwoeck11}
calculations reported in this work were based on the QSpace tensor library
\cite{Wb12_SUN}.  This allowed us to fully exploit the
underlying \SU{2} spin symmetry, as well as to
simultaneously target a range of low lying eigenstates.
Given the simplicity of the model, rungs were considered
as a single site in the DMRG calculations.  This had the
advantage that the $\Jx$ term in Eq.~(1) of the main text can be 
written as a plain nearest-neighbor interaction.  

\subsection{Even vs Odd Ladder Lengths: Uniform \VBSm phase close to $\Jp=0$}

In the main text, we focus on ladders with an even number $L$
of sites along each chain.  This is particularly important for the \VBSm phase which spontaneously breaks translational
symmetry in a valence bond crystal (VBC) like fashion. As
a direct consequence, its local properties are very
sensitive to the specific length of a finite size ladder.
For periodic boundary conditions, the \VBSm phase has a two-fold degenerate g.s.~with an even number of rungs.  For open boundary conditions, a \VBSm ladder with  even $L$ has a unique ground state.  However,
for an odd leg ladder with open boundary conditions, this
picture becomes highly distorted -- in effect, the lowest energy state of such a
ladder would correspond to an excited state of a \VBSm ladder with even length.

A representative DMRG study is shown in \Fig{Fig:VBSm}.
For even $L$ (with $J_\perp =0$), we find a unique g.s. [\Fig{Fig:VBSm}(a)], even in the thermodynamic limit $1/L\to0$.  
Much more remarkably still, at the same $\Jp=0$ as in (a), we observe a degeneracy of the first singlet
and triplet states [\Fig{Fig:VBSm}(d)].   Furthermore, if a small rung coupling $\Jp$
is turned on, the system develops a singlet-triplet gap
whose sign depends on the sign of $\Jp$
[\Fig{Fig:VBSm}(b-c)]!

\subsection{Non-uniform Ladders}

For non-uniform ladders, 
we switch between phases by tuning the parameters 
$J_\perp, J_\times$ in Eq.~(1) of the main text along the ladder,
using the function
\begin{subequations}
\label{eq:transition:func}
\begin{eqnarray}
   f(x)=\frac{1}{1+\exp\left({\tfrac{x}{\delta x}}\right)}
\text{ .}\label{eq:transition:fx}
\end{eqnarray}
This represents a step that is smoothened
over a width $\delta x$.
For a slab geometry A--B--A, with a sandwiched phase
B in the middle of the ladder surrounded by phase A on
either side, we tune the couplings $J_\perp,J_\times$ using the window function
\begin{eqnarray}
  w(x)=f(x-x_+)  - f(x-x_-)
\label{eq:window:wx}
\end{eqnarray}
\end{subequations}
which is non-zero over a stretch $L_c \equiv x_+-x_-$
with $x_\pm=(L\pm L_c)/2$ in the center of the ladder,
and smoothed at the transition points over a width
$\delta x$.

\begin{figure}[t]
\includegraphics[width=0.9\linewidth]{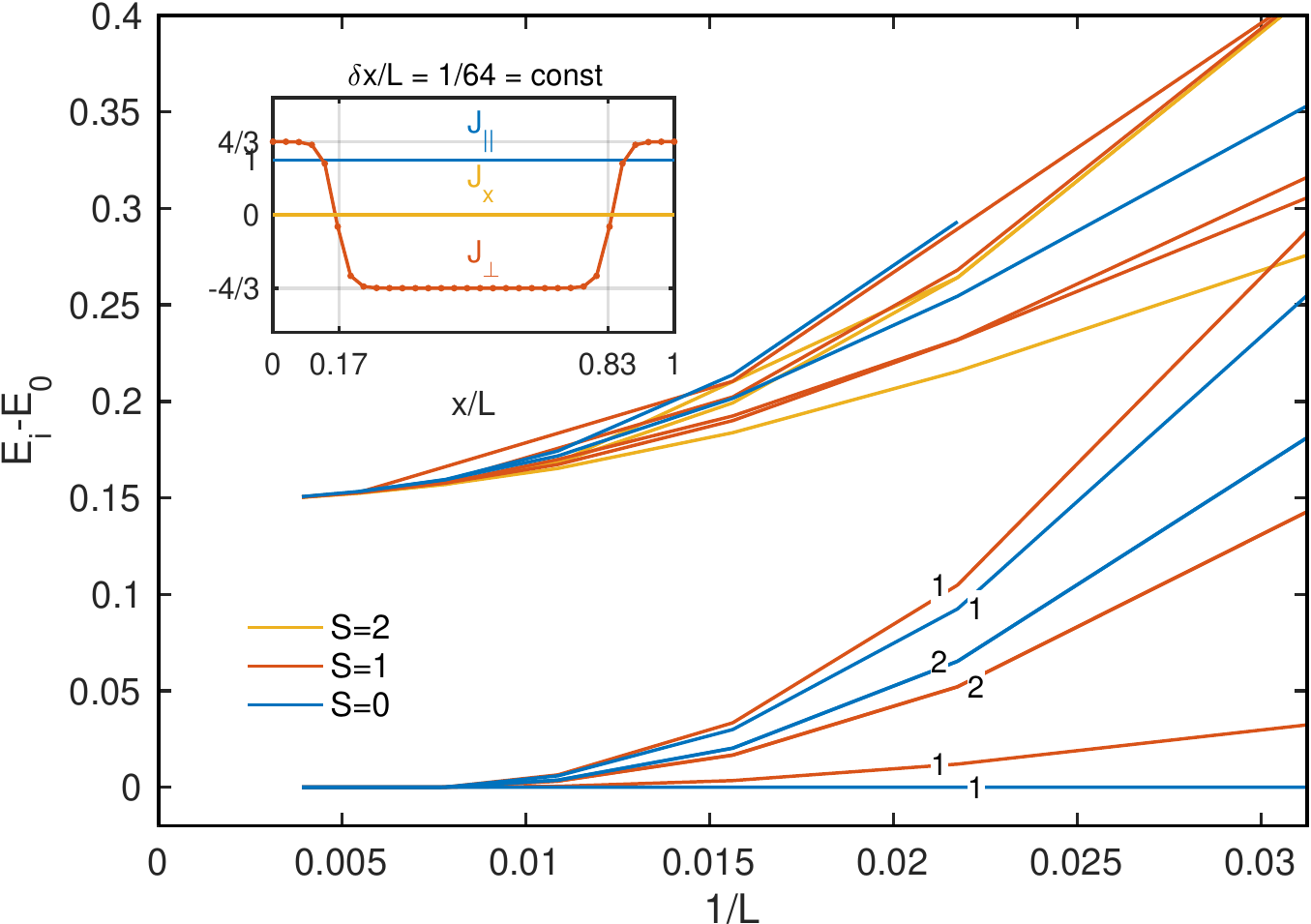}
\caption{
   Development of g.s. degeneracies in an
   RS--H--RS ladder with increasing system size,
   $L=32,\ldots,256$. The DMRG calculations target the lowest $16$
   energy eigenstate multiplets, where lines of the same  color belong to the same global symmetry sector (see
    legend). The width of the boundary between RS
   and H phases was kept constant relative
   to $L$, i.e., $\delta x=L/64$, 
   with the values vs ladder position indicated in the inset for $L=32$.
   As $L\rightarrow\infty$, the ladder develops
   a 16-fold g.s. degeneracy, consisting of four singlets ($S=0$)
   and four triplets ($S=1$), where the numbers on top
   of the lines indicate their degeneracy.  Also shown are
   the first few excited states, illustrating the excitation gap.
}
\label{Fig:DMRG_RSHaldane}
\end{figure}

%%%%
\subsubsection{RS--\VBSm--RS Ladders}
\label{app:RSVBSm}
%%%%

A more detailed analysis of the RS--\VBSm--RS
slab geometry, cf.~Fig. 3 in the
main part, is shown in \Fig{Fig:RS-VBSm-RS-ext}.
Here the size $L_c$ of the central region
is varied w.r.t. to fixed $L_c \mod 2$ in order to analyze even-odd effects of $L_c$ 
for fixed (narrow) transition width $\delta x$
[see legend with panel (c)].
For each system, there is one blue line split off from the 
remainder of the data which thus shows exponential
convergence of a pair of g.s.~singlets
in an otherwise gapped system.
For very small $\delta x$,
the g.s.~doublet
remains split in the thermodynamic limit [(b)].
Yet when going to slightly larger $\delta x$,
rapid convergence towards an exact g.s.~degeneracy is observed [(d)].

%%%%
\subsubsection{\VBSm--RS--\VBSm Ladders}
\label{app:VBSmRSVBSm}
%%%%
One might think that a \VBSm--RS--\VBSm ladder would also exhibit a g.s.~degeneracy dictated by the Jackiw-Rebbi (JR) mechanism because there
is a fermion mass sign change at each of the two boundaries.  However, such a degeneracy is not possible because
of the parity selection rule. In fact, of the two g.s.~candidates,
\begin{equation}
|0_+,0_-;0_+,0_-;0_+,0_-\rangle, \quad |0_+,0_-;0_+,1_-;0_+,0_-\rangle,
\end{equation}
only the first one is allowed.
We demonstrate  in \Fig{Fig:DMRG_VBSmRS} that our DMRG computations are consistent
with this observation.\\

%%%%%%
\subsubsection{RS--Haldane-RS Ladder}
%%%%%%

Finally, we consider the RS--H--RS setup. The arrangement of couplings along this ladder is pictured in the inset of
Fig.~\ref{Fig:DMRG_RSHaldane}.  Keeping the labelling convention 
$|a_+,b_+,c_+;a_-,b_-,c_-\rangle$ for the possible states, the following 16 
states are permitted by parity,
\begin{eqnarray}
|0_+,0_+,0_+;0_-,0_-,0_-\rangle; \quad |0_+,0_+,0_+;0_-,1_-,1_-\rangle;\cr\cr
|0_+,0_+,0_+;1_-,0_-,1_-\rangle; \quad |0_+,0_+,0_+;1_-,1_-,0_-\rangle;\cr\cr
|0_+,0'_+,0_+;0_-,0_-,0_-\rangle; \quad |0_+,0'_+,0_+;0_-,1_-,1_-\rangle;\cr\cr
|0_+,0'_+,0_+;1_-,0_-,1_-\rangle; \quad |0_+,0'_+,0_+;1_-,1_-,0_-\rangle;\cr\cr
|0_+,1_+,0_+;1_-,0_-,0_-\rangle; \quad |0_+,1_+,0_+;0_-,1_-,0_-\rangle;\cr\cr
|0_+,1_+,0_+;0_-,0_-,1_-\rangle; \quad |0_+,1_+,0_+;1_-,1_-,1_-\rangle;\cr\cr
|0_+,1'_+,0_+;1_-,0_-,0_-\rangle; \quad |0_+,1'_+,0_+;0_-,1_-,0_-\rangle;\cr\cr
|0_+,1'_+,0_+;0_-,0_-,1_-\rangle; \quad |0_+,1'_+,0_+;1_-,1_-,1_-\rangle. \nonumber
\end{eqnarray}
Now simply because we can form 16 possible potential ground states consistent with the parity selection rule does not
mean that all will be actually possible.  It could be that there is some ${\cal O}(1)$ energy cost to gluing together
the different phases.  However, in this ladder, all four Majorana fermions change sign at the RS-H boundaries and so the JR mechanism
 (preliminarily) suggests a 16-fold degeneracy.
(An example where we might not expect all allowed states to be ground states is given by  the H-\VBSp-H ladder.
Such a ladder has 320 potential g.s.'s allowed by the parity rule.  However, by JR in combination with the fractionalized spin-1/2's
that sit at the ends of the ladder because of the positioning of the Haldane phase, we actually only expect an 8-fold g.s.~degeneracy.) 
We verify in Fig.~\ref{Fig:DMRG_RSHaldane} from DMRG that indeed the RS--H--RS ladder has a 16-fold degenerate
g.s.  It is decidedly non-intuitive that we can increase the Haldane phase's g.s.~degeneracy by a factor of four merely by
placing it in between two SPT trivial RS phases.
In the spin language it is however relatively straightforward to understand.
Because $J_\perp=0$ at the boundary between phases, we can imagine
a free spin-1/2  at the boundary on both
legs of the ladder which results in a total of $2^4=16$
degenerate states. The two boundary spin-1/2's can be
combined into a singlet and a triplet, hence the systematic
pairing of singlets with triplets.

\bibliography{ladder_bib}
\end{document}